\documentclass[sn-standardnature]{sn-jnl}% Standard Nature Portfolio Reference Style
\usepackage{subfig} 

\jyear{2021}%

\def\aj{Astron.~J.}                   % Astronomical Journal
\def\apj{Astrophys.~J.}                 % Astrophysical Journal
\def\apjl{Astrophys.~J.}                % Astrophysical Journal, Letters
\def\aap{Astron.~Astrophys}                % Astronomy and Astrophysics
\def\araa{Annual Rev.~in Astron.~Astrophys}                % Astronomy and Astrophysics
\def\mnras{Mon. Not. R. Astron. Soc.}             % Monthly Notices of the RAS
\def\pasj{PASJ}               % Publications of the ASJ
\def\pasa{Astron. Soc. Aust.}               % Publications of the ASA
\def\nat{Nature}              % Nature

\newcommand{\mtp}{PSR\,J0901$-$4046}
\def\arcsec{\hbox{$^{\prime\prime}$}}
\newcommand\farcs{\hbox{$.\!\!^{\prime\prime}$}}

\begin{document}

\title[A radio-loud neutron star with an ultra-long spin period]{Discovery of a radio emitting neutron star with an ultra-long spin period of 76 seconds}

%%=============================================================%%
%% Prefix	-> \pfx{Dr}
%% GivenName	-> \fnm{Joergen W.}
%% Particle	-> \spfx{van der} -> surname prefix
%% FamilyName	-> \sur{Ploeg}
%% Suffix	-> \sfx{IV}
%% NatureName	-> \tanm{Poet Laureate} -> Title after name
%% Degrees	-> \dgr{MSc, PhD}
%% \author*[1,2]{\pfx{Dr} \fnm{Joergen W.} \spfx{van der} \sur{Ploeg} \sfx{IV} \tanm{Poet Laureate} 
%%                 \dgr{MSc, PhD}}\email{iauthor@gmail.com}
%%=============================================================%%

\author*[1,2,3]{\fnm{Manisha} \sur{Caleb}}\email{manisha.caleb@manchester.ac.uk}
\equalcont{These authors contributed equally to this work.}

\author*[4,5,6]{\fnm{Ian} \sur{Heywood}}\email{ian.heywood@physics.ox.ac.uk}
\equalcont{These authors contributed equally to this work.}

\author[1,7]{\fnm{Kaustubh} \sur{Rajwade}}
\author[1]{\fnm{Mateusz} \sur{Malenta}}
\author[1]{\fnm{Benjamin} \sur{Stappers}}
\equalcont{These authors contributed equally to this work.}
\author[8]{\fnm{Ewan} \sur{Barr}}
\author[8]{\fnm{Weiwei} \sur{Chen}}
\author[1]{\fnm{Vincent} \sur{Morello}}
\author[1]{\fnm{Sotiris} \sur{Sanidas}}
\author[4]{\fnm{Jakob} \sur{van den Eijnden}}
\author[8,1]{\fnm{Michael} \sur{Kramer}}
\author[9,10,11]{\fnm{David} \sur{Buckley}}
\author[9,10]{\fnm{Jaco} \sur{Brink}}
\author[12]{\fnm{Sara Elisa} \sur{Motta}}
\author[10]{\fnm{Patrick} \sur{Woudt}}
\author[1]{\fnm{Patrick} \sur{Weltevrede}}
\author[1]{\fnm{Fabian} \sur{Jankowski}}
\author[1]{\fnm{Mayuresh} \sur{Surnis}}
\author[6]{\fnm{Sarah} \sur{Buchner}}
\author[1]{\fnm{Mechiel Christiaan} \sur{Bezuidenhout}}
\author[1,13]{\fnm{Laura Nicole} \sur{Driessen}}
\author[4]{\fnm{Rob} \sur{Fender}}

\affil[1]{\orgdiv{Jodrell Bank Centre for Astrophysics, Department of Physics and Astronomy}, \orgname{The University of Manchester}, \orgaddress{\street{Oxford road}, \city{Manchester}, \postcode{M13 9PL}, \country{United Kingdom}}}

\affil[2]{\orgdiv{Sydney Institute for Astronomy, School of Physics}, \orgname{The University of Sydney}, \orgaddress{\city{Sydney}, \postcode{2006}, \state{NSW}, \country{Australia}}}

\affil[3]{\orgname{ASTRO3D: ARC Centre of Excellence for All-sky Astrophysics in 3D}, \orgaddress{\city{Canberra}, \postcode{2601}, \state{ACT}, \country{Australia}}}

\affil[4]{\orgdiv{Astrophysics, Department of Physics}, \orgname{University of Oxford}, \orgaddress{\street{Keble road}, \city{Oxford}, \postcode{OX1 3RH}, \country{United Kingdom}}}

\affil[5]{\orgdiv{Department of Physics and Electronics}, \orgname{Rhodes University}, \orgaddress{\street{PO Box 94, Makhanda}, \city{Grahamstown}, \postcode{6140}, \country{South Africa}}}

\affil[6]{\orgname{South African Radio Astronomy Observatory}, \orgaddress{\street{Fir Street, Black River Park}, \city{Cape Town}, \postcode{7925}, \country{South Africa}}}

\affil[7]{\orgname{ASTRON, the Netherlands Institute for Radio Astronomy}, \orgaddress{ \street{Oude Hoogeveensedijk 4}, \postcode{7991 PD}, \city{Dwingeloo}, \country{The Netherlands}}}

\affil[8]{\orgname{Max-Planck-Institut f{\"u}r Radioastronomie}, \orgaddress{\street{D-53121}, \city{Bonn}, \postcode{53121}, \country{Germany}}}

\affil[9]{\orgname{South African Astronomical Observatory}, \orgaddress{\street{PO Box 9, Observatory Road}, \city{Cape Town}, \postcode{7935}, \country{South Africa}}}

\affil[10]{\orgdiv{Department of Astronomy}, \orgname{University of Cape Town}, \orgaddress{\street{Private Bag X3, Rondebosch}, \city{Cape Town}, \postcode{7701}, \country{South Africa}}}

\affil[11]{\orgdiv{Department of Astronomy}, \orgname{University of the Free State}, \orgaddress{\street{PO Box 339, Bloemfontein}, \city{Cape Town}, \postcode{9300}, \country{South Africa}}}

\affil[12]{\orgname{INAF–Osservatorio Astronomico di Brera}, \orgaddress{\street{via E. Bianchi 46}, \city{Merate (LC)}, \postcode{23807}, \country{Italy}}}

\affil[13]{\orgname{CSIRO, Space and Astronomy}, \orgaddress{\street{P.O. Box 1130}, \city{Bentley}, \postcode{3115}, \state{WA}, \country{Australia}}}

%%==================================%%
%% sample for unstructured abstract %%
%%==================================%%

% Abstract – up to 150 words, unreferenced.

\abstract{The radio-emitting neutron star population encompasses objects with spin periods ranging from milliseconds to tens of seconds. As they age and spin more slowly, their radio emission is expected to cease. We present the discovery of an ultra-long period radio-emitting neutron star, \mtp\,, with spin properties distinct from the known spin and magnetic-decay powered neutron stars. With a spin-period of 75.88 s, a characteristic age of 5.3 Myr, and a narrow pulse duty-cycle, it is uncertain how radio emission is generated and challenges our current understanding of how these systems evolve. The radio emission has unique spectro-temporal properties such as quasi-periodicity and partial nulling that provide important clues to the emission mechanism. Detecting similar sources is observationally challenging, which implies a larger undetected population. Our discovery establishes the existence of ultra-long period neutron stars, suggesting a possible connection to the evolution of highly magnetized neutron stars, ultra-long period magnetars, and fast radio bursts.}

% \keywords{keyword1, Keyword2, Keyword3, Keyword4}

\maketitle

\section{Introduction}\label{sec1}

Radio pulsars are rotation-powered neutron stars, which emit coherent beams of radio emission generated by highly relativistic particles in regions above their magnetic poles. Their known spin periods ($P$) range from 1.4\,ms to 23.5\,s and they are divided into various sub-classes (e.g. rotating radio transients, millisecond pulsars, magnetars - \url{https://www.atnf.csiro.au/research/pulsar/psrcat/}) depending on their observational properties. Particle acceleration and abundant electron–positron pair production is postulated to be an essential condition for the coherent radio emission from pulsars, with the particle acceleration potential expected to be lower for larger spin periods. As seen in most neutron stars, the radio emission is also expected to be strongly inhibited, or cease if the magnetic field configuration and strength, exceed the quantum critical field ($B_\mathrm{cr} = 4.413 \times 10^{13}$\,G) \citep{BH98}. Here, we present the discovery of a highly-magnetized 75.88\,s period radio-emitting neutron star, \mtp\, which challenges these conditions for, and the nature of, the radio emission and raises questions about the spin evolution of neutron stars in general.

\section{The Data}
\subsection{The discovery and properties of \mtp}\label{sec2}

\mtp\, was a serendipitous single pulse discovery at 1284 MHz on 27 September 2020, in an observation directed at the high mass X-ray binary, Vela X-1, during simultaneous image and time domain searches by the Meer(more) TRAnsients and Pulsars (MeerTRAP - \url{https://www.meertrap.org/}) and ThunderKAT (\url{http://www.thunderkat.uct.ac.za}) projects at the MeerKAT radio telescope in South Africa. The pulse was initially detected in the MeerTRAP beamformed data in a single coherent tied-array beam of angular diameter $\sim45$ arcseconds. A review of the MeerTRAP data for that observation revealed that there were further wide, but weaker pulses, which were missed by the real-time single pulse detection system. A total of fourteen pulses were identified in the beamformed time domain searches, which were regularly spaced in a span of $\sim 30$ minutes. A periodicity analysis resulted in an initial period of $P = 75.89 \pm 0.01$ seconds. The corresponding full time and frequency integration image of the field revealed an associated point source at the location of the coherent beam. These data were re-imaged at the smallest possible integration time of 8 seconds and more pulses were identified. An initial inspection of the 8-second images from two other epochs where MeerTRAP data were not available, also revealed that the source exhibited a consistent periodicity. These snapshot images allowed the source to be localized to arcsecond precision. The deepest image of the field shows a partially visible, diffuse shell-like structure surrounding \mtp\,, which is possibly the supernova remnant from the event that formed the neutron star. The complexity of the field in terms of diffuse emission requires additional analysis to determine a robust association of this radio shell with \mtp. No known pulsars are located within 2 degrees of this sky location.

A total of six L-band (856 -- 1712 MHz) and one UHF-band (544 -- 1088 MHz) observations have been performed between September 2020 and May 2021. During these, we detect single pulses from every rotation of the source. The L-band data have resulted in the timing solution shown in Table \ref{tab:timingparams}. Despite the large jitter in the pulse shapes of single pulses, we obtain remarkably stable pulse profiles over the various epochs due to the high signal-to-noise ratios (S/Ns). Using 29 times of arrival (ToAs),
% define this acronym?
typically two per epoch, over 7.4 months, we measure timing residuals with a low root-mean-square (rms) of 5.7~ms (see Extended Data Figure \ref{fig:timing}). When compared to the pulse period, the fractional accuracy of $\sim7 \times 10^{-5}$ is comparable to the most accurately timed millisecond pulsars. We do not find any evidence of timing noise or covariance of spin parameters with position. \mtp\, has a best fit dispersion measure (DM) of $52 \pm 1$ pc\,cm$^{-3}$ and average half-power pulse widths of $\sim300$\,ms at both L- and UHF-band suggesting no evidence for radius-to-frequency mapping.
We measure pulse-averaged peak flux densities of 89.3~$\pm$~2.7 mJy beam$^{-1}$ and 169.3~$\pm$~14 mJy beam$^{-1}$ at L-band and UHF-band, respectively, with a period-averaged flux density of $408 \pm 5\,\mu$Jy beam$^{-1}$ at L-band.
The measured DM corresponds to distances of approximately 0.3 and 0.5~kpc according to the \textsc{ymw16} \citep{ymw16} and \textsc{ne2001} \citep{ne2001} Galactic electron density models, respectively. The period ($P = 75.88$~s) and period derivative ($\dot{P} = 2.25 \times 10^{-13}$~s s$^{-1}$; pulsar spin-down rate) correspond to a characteristic age, surface magnetic field strength, and spin-down luminosity of 5.3~Myr, $1.3\times 10^{14}$~G and $2.0\times10^{28}$~erg~$\mathrm{s}^{-1}$ assuming a dipolar magnetic field configuration, respectively (see Figure \ref{fig:ppdot}). This discovery confirms the existence of ultra-long period neutron stars.

\section{Results}
\subsection{Radio emission properties}\label{sec3}

Single pulse analyses of the radio emission from \mtp\, reveal remarkable and unusual spectro-tempo-polarimetric properties, quite unlike anything seen in known radio pulsars. We notice that the pulse shape is variable both inter-epoch and intra-epoch, but some features persist. Overall, the single pulses studied over 6 epochs can be grouped into 7 different types, namely: normal, quasi-periodic, spiky, double-peaked, partially nulling, split-peak and triple-peaked as shown in Figure \ref{fig:morphology}. Although magnetars are sometimes seen to emit wide, bright radio pulses that comprise several sub-pulse components of varying widths and amplitudes, these are more chaotic within and between subsequent pulses.  

In some of the bright pulses we measure a quasi-periodicity in the sub-pulse components which at times appear to be harmonically related between pulses (see Extended Data Figure \ref{fig:acf}). In some others we see multiple quasi-periods within a single rotation as seen in Extended Data Figure \ref{fig:multiperiodACF}. Overall, the quasi-periods are common across the UHF- and L-band observations. We observe the width of the sub-pulse components in \mtp\, to be exactly half of the quasi-period. The shortest and longest quasi-periods we measure are 9.57~ms (104~Hz) and 338~ms (2.96~Hz) respectively (see Extended Data Figure \ref{fig:QP}). Similar quasi-periodic features have been observed in fast radio bursts (FRBs) \citep{chime_QP}. Radio observations of the magnetar XTE~J1810$-$197 following its 2018 outburst revealed a persistent 50-ms periodicity imprinted on the pulse profile \citep{lld+19}. 
The most commonly seen quasi-period across all observations is $\sim76 $~ms (13~Hz), which is $\approx P/1000$.
This quasi-period follows the spin-period scaling seen in corresponding values of the micropulses in normal pulsars \citep{cor79}. This scaling can be most easily associated with the emission of beamlets making up the wider sub-pulses \citep{kjv02}, suggesting that the periodicities are caused by a {\em temporal} or {\em angular} mechanism rather than the motion of the beamlets in the polar cap region. Alternatively, this quasi-period could be related to sub-pulses or drifting sub-pulses. Each of the sub-pulses or dense, isolated `sparks' (i.e. pair-production sites) are theorized to have a corresponding plasma column, which radiates and generates the observed sub-pulses, which may rotate around the magnetic axis. Such quasi-periodic oscillations are also theorized in models of FRBs, where they are due to magneto-elastic axial (torsional) crustal eigenmodes originating close to the neutron star surface \citep{WC20}. The eigenfrequencies of these oscillations are expected to depend most strongly on the neutron star mass and the crust equation-of-state \citep{WC20}. These local crustal oscillations can create Alfv\'en waves that propagate to larger heights in the magnetosphere, thereby producing an oscillating $E_{||}$ in the charge starved region to produce the observed coherent radio emission \citep{LKZ20}. 

Ultimately, it is unclear what causes the quasi-periodicity in \mtp. Global magnetoelastic axial (torsional) oscillations are a tempting explanation, but the persistence of our periodicities would require repeated triggers and/or very long damping times. The observed periodicities and frequencies, however, may be consistent with models proposed for magnetars, and the similarity with the periodic feature of the radio-loud magnetar XTE~J1810$-$197 are intriguing. We note that \mtp's position in the $P-\dot{P}$ parameter space is offset from the known magnetar population. We also note that \mtp\, may differ in other physical quantities (such as in its mass) which we cannot access from  our observations but which are likely to play a role in the seismic properties of neutron stars. Hence, differences in the behaviour compared to other neutron stars or magnetars may not be unexpected. It has been proposed \citep{BW+20} that bright coherent radio bursts can be produced by highly magnetized neutron stars that have attained long rotation periods (few 10s to a few 1000s of seconds) called Ultra-Long Period Magnetars (ULPMs). Recently, a source GLEAM-X~J$162759.5-523504.3$ with a period of $\sim20$~minutes in the radio has been discovered, and is speculated to be a member of this class \citep{hzb+22}. X-ray isolated neutron stars (XINS) are nearby cooling neutron stars with spin periods in the range $3.4–11.3$~s \citep{YHN19} and are characterized by thermal, soft X-ray, emission. They are believed to be old, strongly magnetized neutron stars despite their non-detection in the radio so far \citep{vrp+13}. A few XINS lie above the low-twist death line in Figure \ref{fig:ppdot}, implying possible ULPM origins. Interestingly, \mtp\, also falls in the parameter space (see Figure \ref{fig:ppdot}) where these ULPMs are expected to exist. \mtp\, could potentially be an old magnetar or a member of the ultra-long period magnetars, a result that needs to be confirmed with future multi-wavelength observations. \mtp\, is therefore an important piece in the puzzle of the evolution of highly magnetized neutron stars and their connection to FRBs.

Typically, when magnetars are radio active, there is also often X-ray emission. We therefore observed \mtp\, in the X-rays using \textit{Swift}/XRT simultaneously with the MeerKAT observations on 2021-02-01 and 2021-02-02 and did not detect any X-ray emission. Assuming a blackbody spectrum with temperature 1.5 keV, and an equivalent column density of  $N_{\rm H}$ = 4.32$\times$10$^{21}$ cm$^{-3}$, we place 3-sigma upper limits of $L_{\rm X1} (2-10 \, \rm{keV})$ $<$ 1.6$\times$10$^{30}$ erg s$^{-1}$ and $L_{\rm X2} (2-10 \, \rm{keV})$ $<$ 3.2$\times$10$^{30}$ erg s$^{-1}$ on the X-ray luminosity for distances $d_1$ $\approx$ 0.3~kpc and $d_2$ $\approx$ 0.5~kpc, respectively. The location of \mtp\, in the $P-\dot{P}$ parameter space is consistent with it having spun-down from a magnetar-like period of 10~s in $\sim5$~Myr, assuming a braking index of 3. However, we do not find any evidence for radical changes in the $\dot{P}$ as seen in most magnetars in the 7.4 months since discovery. Additionally, while magnetars are observed to have shallow radio spectra (e.g. \citep{Kramer1810, DLB+19}), \mtp\, has a measured L-band in-band spectral index of $-1.7\pm0.9$, which is more consistent with the pulsar population.
Canonical, rotation powered pulsars are observed to have X-ray luminosities much smaller than their spin-down luminosities, with on average $L_{\rm X} \approx 10^{-3}\dot{E}$ \citep{BT1997}. Conversely, magnetars are seen to have $L_{\rm X} \gtrsim \dot{E}$. For \mtp\,, based on the X-ray luminosity upper limit and the spin-down $\dot{E}$ in Table \ref{tab:timingparams}, we see $L_{\rm X} \lesssim 10^{2} \dot{E}$. This places it closer to magnetars but is not constraining. Additionally, the single pulse brightness is seen to vary significantly in the 8,726 2-second integration time images, across the 6 L-band and 1 UHF-band epochs. The source appears to have secularly grown fainter (see Extended Data Figure \ref{fig:lightcurves} and Table \ref{tab:observations}), from a mean pulse brightness of 16.4~$\pm$~7.9~mJy beam$^{-1}$ for the observations centered on 59246.087481292554 to 12.9~$\pm$~5.2~mJy beam$^{-1}$ on 59343.62301600376, suggesting a dynamic magnetosphere transforming on timescales much faster than associated with the characteristic age, $\tau$. If this is indeed part of a long-term dimming of the source, and not a short-term variation, then it is also reminiscent of radio-loud magnetars transitioning into quiescence. 

The single pulse polarization profiles of \mtp\, show complex structure, and on average are more circularly than linearly polarized (see Extended Data Figure \ref{fig:polnprof}). This is not unexpected in radio-loud neutron stars, particularly magnetars. The magnetar J1622$-49$60 exhibits different categories of pulses of varying polarization fractions. One particular category shows a higher value of circular polarization. The Faraday rotation measure (RM) towards \mtp\, is measured to be $-64 \pm 2$ rad~m$^{-2}$. The RM of \mtp\, is consistent with the contribution from the smoothed Galactic foreground \citep{OJG+15} and with the RMs of nearby pulsars. This therefore precludes the presence of a significant intrinsic RM imparted at the source. A phase resolved histogram of the polarization position angle shows the characteristic S-shaped curve expected from a rotating magnetic dipole (see Supplementary Figure 3). This suggests that the line-of-sight passes close to the magnetic pole as we see the S-shaped curve even within a 1\% duty cycle. This is consistent with our constraint on the impact parameter of $\beta \lesssim 0.2^{\circ}$, using a rotating vector model fit.

\section{Discussion}

The \mtp\, pulses classified as split-peak are the most common, $\sim33\%$ of all pulses across all observations, closely followed by a combination of the quasi-periodic and partially nulling pulses which together form $\sim34\%$. The normal and spiky pulses comprise $\sim27\%$ and $\sim6\%$ respectively. A comparison of the energies of the various pulse shape archetypes shows that despite the enormous variability seen in the pulse profile shapes, their energies span more or less the same range (see Supplementary Figure 8). For instance, we lose $\sim40\%$ of the energy to the dropouts/dips seen in the quasi-periodic and partially nulling pulses, which when accounted for by modelling the pulse envelope, is similar to the energy distribution of the `split-peak' and possibly also the `normal' pulses. This suggests that the pulses with dropouts/dips are not drastically brighter than the other types, implying that an overall increase in particle flow cannot be responsible.

\begin{figure}
    \centering
    \includegraphics[width=4.5 in]{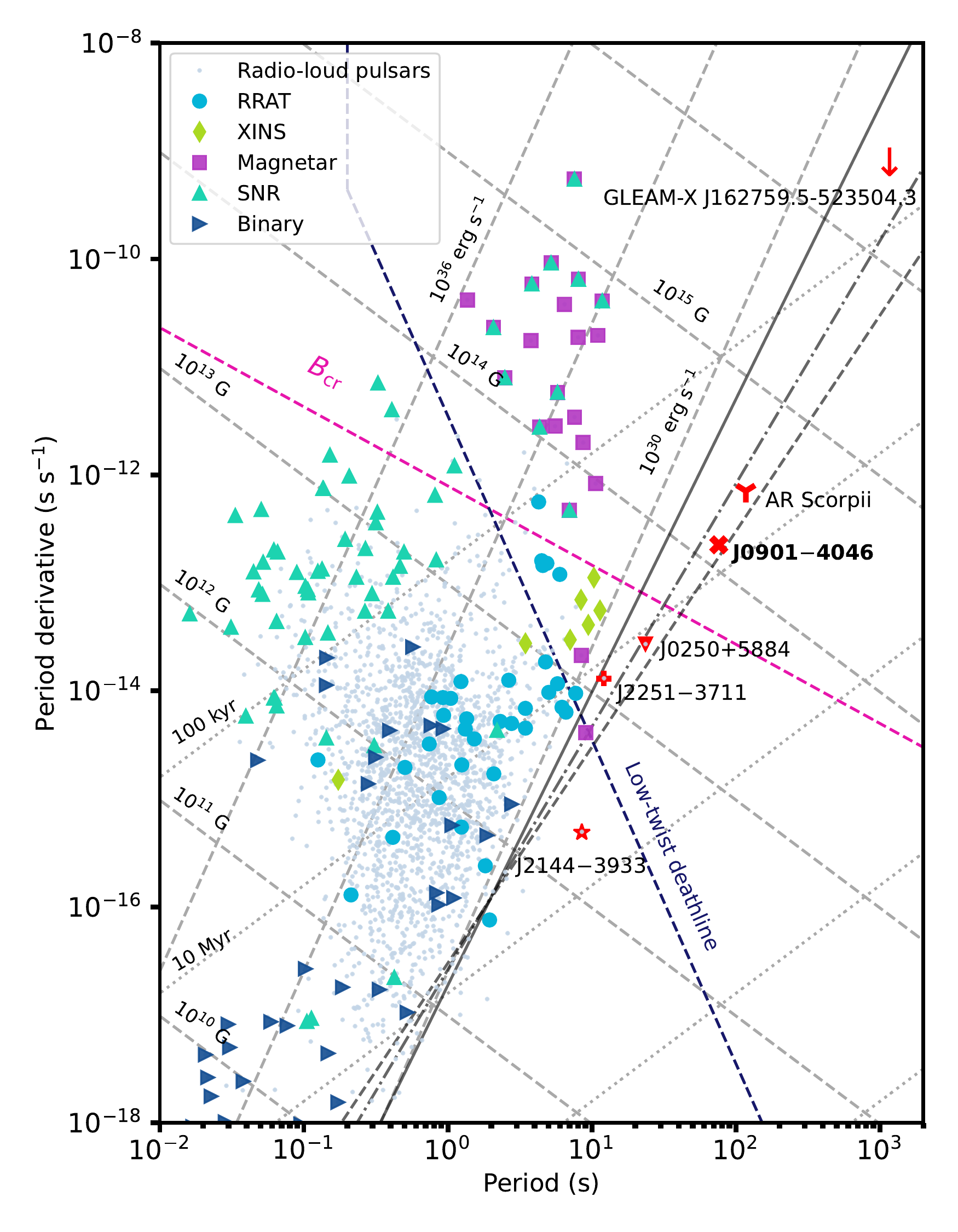}
    \caption{$P-\dot{P}$ diagram based on the ATNF pulsar catalog. The various sub-classes of pulsars are represented by the markers in the legend. The longest spin period radio pulsars and the white dwarf binary system AR Sco are highlighted in red. Lines of constant age and magnetic field are shown as dotted and dashed lines respectively. The lower right corner of the figure represents the `death valley' with various death lines from the literature, where sources below these lines are not expected to emit in the radio. The solid death line represents Equation 9 in CR93 \citep{CR93}. In dot-dashed and dashed are the death lines modeled on curvature radiation from the vacuum gap and SCLF models as shown by Equations 4 and 9 respectively in \citep{ZHM2000}. Sources above the low-twist death line are potential ultra-long period magnetars.}
    \label{fig:ppdot}
\end{figure}

\begin{figure}
\centering
  \includegraphics[width=5.5 in]{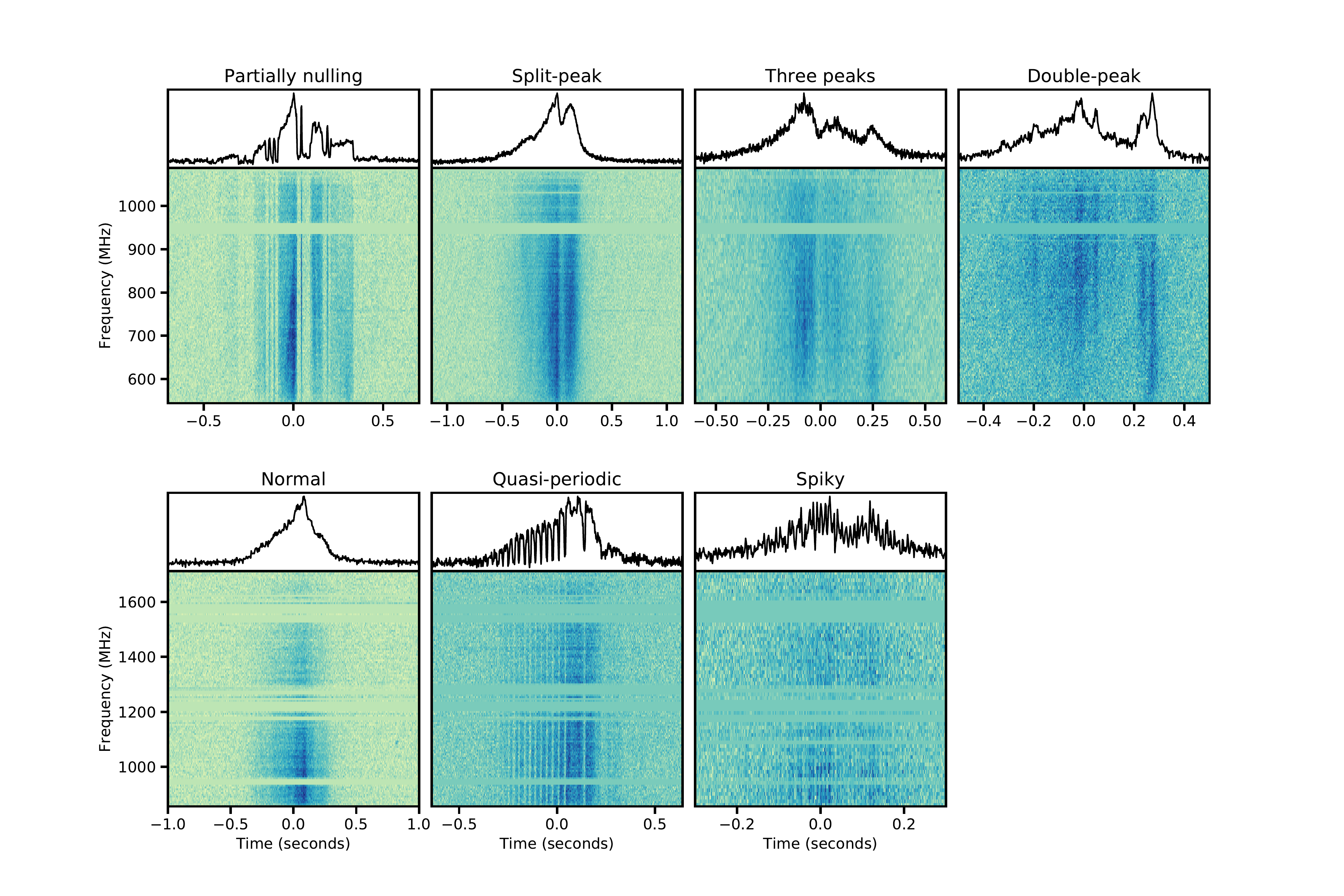}
    \caption{Gallery of the pulse morphology types of \mtp\,. The morphological type is given in the title for each panel. The top panels are pulses observed in the UHF-band while the bottom panels are pulses observed at L-Band.}
\label{fig:morphology}
\end{figure}

The measured period implies an extremely large pulsar light-cylinder (of radius $R_\mathrm{LC} = cP/2\pi = 3.62 \times 10^{6}$\,km) and consequently a relatively compact polar cap (with radius $R_\mathrm{p} = \sqrt{2\pi R^3/cP} = 16.62$\,m, where $R = 10$\,km). For an assumed emission height of hundreds of kilometers above the surface, the beam width, and consequently the duty cycle of \mtp\, is small ($\sim$1\%) and found to be consistent with the empirical scaling relation between pulse width and spin period ($W \propto P^{-0.5}$) observed in canonical pulsars (e.g. \citep{JK2019}). 
\mtp\, is seen to lie in the $P-\dot{P}$ pulsar parameter phase space far from the other recently discovered slowest spinning radio emitting pulsars with periods of 23.5\,s \citep{TBC+18} and 12.1\,s \citep{mke+20} respectively. It is also located beyond the `death line' as defined by the RS75 \citep{RS75} and CR93 \citep{CR93} inner vacuum-gap (IVG) curvature radiation models for radio emission (see Figure \ref{fig:ppdot}). These models suggest pulsars in this region cannot support the pair-cascade production just above the pulsar polar cap in their inner magnetospheres that is required to sustain the observed radio emission. This is because at large spin periods, it is no longer possible to achieve the increase in thickness of the vacuum gap above the neutron-star polar cap needed to maintain the required potential difference for pair production. This leads to the cessation of radio emission. However, \mtp\, does lie above the space-charge-limited flow (SCLF) radio emission model death line where pair cascade can be supported through non-relativistic charges flowing freely from the polar cap if there is a multipolar magnetic field configuration. Unambiguous signatures of the presence of multipolar components close to the neutron star surface have been seen in magnetars \citep[SGR~$0418+5729$;][]{tem+13}, and more recently in an accreting millisecond pulsar \citep[PSR J$0030+0045$;][]{rwb+19, rrw+19} suggesting a likely ubiquity of a multipolar magnetic field configuration in neutron stars.

The putative boundary for radio quiescence in Figure \ref{fig:ppdot} indicated by $B_{\rm cr}$ lies about an order of magnitude below the position of \mtp. The quantum process of single photon pair production ($\gamma \rightarrow e^{+} e^{-}$) is expected to dominate below the $B_{\rm cr}$ line resulting in predominantly `radio loud' pulsars. The quantum process of photon splitting ($\gamma \rightarrow \gamma \gamma$) is expected to dominate above the $B_{\rm cr}$ line resulting largely in `radio quiet' pulsars due to the suppression of pair creation. \mtp\, lies above, and at a similar distance from the $B_{\rm cr}$ line as many of the magnetars. Unlike magnetars the radio emission of \mtp\, has a small duty cycle, but like the magnetars it is highly variable. High-B radio pulsars have on occasion been observed to exhibit magnetar-like activity and have been termed `quiescent magnetars' \citep{kb17}. Radio emission from magnetars is usually transient and often follows a high-energy outburst (e.g. \citep{CRH+06}). It is therefore useful to see how long this source has been a radio emitter and if any previous unidentified high-energy transient has been seen in this region. We did not find any historical high-energy transients coincident with the location of \mtp. Unfortunately, none of the relevant radio continuum surveys, TIFR GMRT Sky Survey (TGSS) \citep{IJM+17}, Sydney University Molonglo Sky Survey (SUMSS) \citep{MMB+03} or the Rapid ASKAP Continuum Survey (RACS) \citep{mcconnell20}, were sensitive enough to detect the source given its current time-averaged flux of a couple of hundred micro-Janskys.
Analyses of the Parkes Multibeam Pulsar Survey (PMPS) data \citep{mlc+01} from nearby pointings also did not detect the source. The nearest pointing should have been sensitive enough, but a combination of radio frequency interference and the hardware high-pass filter likely prevented a detection. 

The discovery of a $\sim 117$~s and 118~s periodicities in the multi-wavelength (including radio) brightness changes of AR Scorpii (AR Sco; a radio pulsating white dwarf binary system) \citep{2016Natur.537..374M} was interpreted as dipole emission from a spinning down of a magnetic white dwarf and not as a neutron star \citep{2017NatAs...1E..29B, ggl+20}. Given the similarity in period to \mtp\ we therefore searched for multi-wavelength counterparts in archival data to determine whether it could be a related system. We identified a 17th mag Gaia source, offset by $\sim1\arcsec$ in right ascension and $\sim3\arcsec$ in declination from the radio coordinates, as a possible optical counterpart. Initial follow-up photometry with the South African Astronomical Observatory (SAAO) 1-m telescope showed indications of long term variability in the star. However, spectroscopic observations with the Southern African Large Telescope (SALT) revealed the optical source to be an A-type star, with narrow Balmer absorption lines. As we see no evidence for hydrogen or helium emission lines and no distinct secondary star component in the spectrum, we rule out the possibility of \mtp\, being an AR~Sco type system, or associated with this A-type star. There are no other obvious counterparts brighter than $20-21$ mag in this region. While the spin-period of \mtp\, might be consistent with a white dwarf, we do not see any multi-wavelength support for this. 

To ascertain if there is an un-pulsed radio component which might be attributed to a pulsar wind nebula, or perhaps indicate emission of a non-neutron star origin, we imaged follow-up MeerKAT visibility data that was recorded at a higher time resolution of 2 seconds. After removing the epochs that contain pulsed emission we obtain a 3$\sigma$ upper limit on the peak brightness of persistent radio emission of 18 $\mu$Jy beam$^{-1}$ at 1284 MHz (see Extended Data Figure \ref{fig:OnOff}). We also find no evidence for pulsed or continuum emission outside of the narrow pulse window, but the radio shell is still present. The extreme ratio of the peak on-pulse flux to the off-pulse flux, the large first period derivative, the timing properties, and the lack of evidence for detections at other wavelengths supports our hypothesis that \mtp\, is a radio emitting neutron star with one of the longest known periods.

\section{Implications for the population of radio-emitting neutron stars}\label{sec5}

Although modern pulsar surveys are sensitive to a wide range of radio emitting neutron stars, the serendipitous discovery of \mtp\, has revealed some of the biases that still remain, and highlights that there may be many more sources like this to be found. The long duration of the pulse, low DM and long period are problematic for commonly used single pulse and periodic pulsar search techniques. 
The very narrow duty cycle suggests a strong bias where many other similar systems may have beams missing the Earth completely.
This suggest that there are many more neutron stars in the Galaxy than the known population suggests, unless many pulsars continue to emit for longer than previously thought or if there is an evolutionary link to another class of neutron star such as the magnetars \citep{KK08}, or perhaps a combination of all of these.
The position of \mtp\, in the $P-\dot{P}$ parameter space along with the unusual single pulse properties such as quasi-periodicity and partial-nulling, make it a potentially very useful target for understanding the radio emission properties of neutron stars across the population. Future image and time domain searches for similar long-period objects could prove vital to our understanding of the Galactic neutron star population and potentially links to FRBs.

\clearpage

\section*{Methods}\label{sec6}

\section*{Calibration of interferometric imaging data}

The MeerKAT observations of the field around \mtp\, are summarised in Table \ref{tab:observations}. Nine distinct observations were used, eight of which used MeerKAT's L-band (856 -- 1712 MHz) receivers, and one of which was observed at UHF band (580 -- 1015 MHz). Total on-target times (T$_{\mathrm{obs}}$) and correlator integration times per visibility point (T$_{\mathrm{int}}$) are listed in Table \ref{tab:observations}. The latter is what limits the time resolution of any individual image of the field. The `discovery' observations, associated with observations of Vela X-1 from the ThunderKAT project, were taken with the correlator configured to deliver 32,768 spectral channels. The follow-up (DDT) observations that targeted \mtp\, directly used 4,096 channels. For the imaging data the observations were in all cases averaged down to 1,024 channels prior to the commencement of the processing. 

The approach to imaging the field was common for all observations. Each observation contains 5 minutes scans of the standard primary calibrator source J0408$-$6465, and the scans of the target field were bracketed by observations of the nearby secondary calibrator J0825$-$5010, which was observed for 2 minutes for every 15 minutes on the target for the ThunderKAT data, and for every 30 minutes for the DDT observations. The calibrator scans were flagged in order to remove radio frequency interference, and the low-gain edges of the telescope's bandpass response. Bandpass, delay, and flux-scale corrections were derived from the observations of the primary calibrator, and time-dependent complex gain and delay corrections were derived from the scans of the secondary. These corrections were then applied to the target data. These steps were all performed using the {\sc casa} package \citep{mcmullin07}.

Following the application of the referenced calibration, the target data were flagged using the {\sc tricolour} (\url{https://github.com/ska-sa/tricolour/}) software. The target field was imaged using {\sc wsclean} \citep{offringa14}. Deconvolution was allowed to proceed in an unconstrained fashion. The field exhibits some complex radio morphology, thus a cleaning mask was derived from the first image, after which the imaging was repeated with deconvolution only proceeding within the masked regions. The frequency dependence of the sky was captured by imaging the data in eight separate sub-bands, using a fourth-order polynomial fit to capture spectral curvature, mainly an instrumentally-induced property due to the frequency dependent antenna primary beam response and the broad bandwidth. A sky area (3.12 $\times$ 3.12 deg$^{2}$) much larger than the main lobe of the primary beam ($\sim$1 deg at FWHM) was imaged in order to deconvolve bright off-axis sources that are detected through the primary beam sidelobes. The use of the cleaning mask, spectral settings, and large sky area in this second imaging run are all to ensure a reliable model for subsequent self-calibration, which consisted of solving for instrumental phase and delay corrections for every 32 seconds of data using the {\sc cubical} package \citep{kenyon18}. The scripts used to perform the data reduction process also provide an exhaustive list of the calibration and imaging parameters, and can be found online (\url{https://github.com/IanHeywood/oxkat} v0.2) \citep{heywood20}.

\section*{Snapshot imaging of \mtp\,}
\label{sec:snapshot-imaging}

To expedite the production of per-integration time images, we first subtract a model of the sky that captures most of the bright emission, but critically does not include any clean components that are associated with \mtp\, itself. For each MeerKAT observation, the self-calibrated data were imaged, the resulting model images were masked at the position of \mtp\,, and the modified model was inverted into a set of model visibilities, which were then subtracted from the data. Images could then be made for each correlator dump time (8 or 2 seconds) to search for pulsed emission from the target. In the case of the ThunderKAT data the visibilities were first phase-rotated to the position of \mtp. Since the dominant emission in the field has been subtracted, small images around the target are viable, and no deconvolution needs to be performed under the (valid) assumption that \mtp\, is unresolved by MeerKAT. This speeds up the imaging process considerably.

Extended Data Figure \ref{fig:lightcurves} shows the peak brightnesses of the pulses as detected in 8,726 2-second snapshot images from the six L-band epochs with this integration time. The pulse brightness varies significantly, however no pulses are missed at the sensitivity limit of our observations, with the exception of the cyan region in the lower left panel, where data were lost due to lightning. The mean RMS noise the snapshot images being 350 $\mu$Jy beam$^{-1}$ with a standard deviation of 50 $\mu$Jy beam$^{-1}$. The right hand column of panels in Extended Data Figure \ref{fig:lightcurves} shows the pulse brightness expressed as a signal-to-noise (S/N) ratio. The brightness is measured as the peak pixel value in a 400 pixel box centered at the position of \mtp\,. The noise is taken to be the standard deviation of the pixels in an off-source box of equivalent size. The blue curve on the right hand column of panels shows the S/N ratio of the peak pixel in the off-axis box. 
\section*{A search for persistent (off-pulse) radio emission}

Jointly imaging and deconvolving the visibilities used to produce the 2-second images results in the image shown in the left panel of Extended Data Figure \ref{fig:OnOff}. The accumulated pulse emission results in the prominent compact source in the centre of the image, with a peak brightness of 40 ($\pm$5.2) $\mu$Jy beam$^{-1}$. Identifying the timestamps of the pulses shown in Extended Data Figure \ref{fig:lightcurves} allows us to re-image the data with those integration times excluded in order to search for off-pulse (persistent) radio emission associated with \mtp\,. This process results in the image shown on the right hand panel of Extended Data Figure \ref{fig:OnOff}. There is a 4.3 $\mu$Jy beam$^{-1}$ ($\sim$1$\sigma$; $\sigma$~=~4.7 $\mu$Jy beam$^{-1}$) peak in the pulse-subtracted radio map spatially coincident with the peak of the pulsed emission in the image formed from the full dataset. We can thus place a 3$\sigma$ upper limit on the peak brightness of a persistent radio source coincident with the peak of the pulsed emission of 18 $\mu$Jy beam$^{-1}$. The other sources visible in Extended Data Figure \ref{fig:OnOff} are faint compact sources that were not deconvolved in the per-epoch imaging process, and thus not subtracted from the visibilities.

In the deepest image we have made in Extended Data Figure \ref{fig:OnOff}, there is diffuse emission in the region of \mtp. More analysis is needed to see if it is somehow associated with \mtp\, as this is a complex region of the sky with lots of diffuse emission. However if it were attributable to radio emission from a supernova remnant that could be associated with \mtp\, it would suggest that the source was much younger than the characteristic age and have important implications for its evolution. None of the features visible in Extended Data Figure \ref{fig:OnOff} are above the noise floor of the 2 second images, and thus do not contaminate the measurements presented in Extended Data Figure \ref{fig:lightcurves}.

\section*{DM estimate}

The DM of each single pulse was estimated by maximizing for structure within the burst envelope using \texttt{DM\_phase} (\url{https://github.com/danielemichilli/DM\_phase}). The structure-optimized DM is determined by maximizing the coherent power across the bandwidth \citep{SMP19}. We de-dispersed the data over a trial DM range of $49.0 \leq \rm{DM} \leq 54.0$ pc cm$^{-3}$ in steps of 0.1 pc cm$^{-3}$. The uncertainty on each DM estimate was calculated by converting the standard deviation of the coherent power spectrum into a standard deviation in DM via the Taylor series. We measure a weighted average DM of $52 \pm 1$ pc\,cm$^{-3}$ for \mtp.

\section*{Timing}

The MeerTRAP pipeline searches data in real time and for each transient event detected, it writes out a short \textsc{sigproc} filterbank file that contains a few seconds of the original input data stream centered around the detection time of the associated event. For each detection, a substantially smaller, second-stage candidate file is also created as follows: the native resolution filterbank file is dedispersed at the detection DM reported by the search pipeline, a reduced time span of the data window equal to the dispersion delay of the candidate DM is extracted and lastly the time and frequency resolution of the data are reduced to an appropriate level according to the reported pulse width for the event (larger widths correspond to a larger acceptable degradation factor of the data). Second-stage candidate files are small enough to be stored \textit{en masse}, but the native resolution filterbank files are not; only those deemed very likely to contain a genuine astrophysical event by an automated classifier are kept, the rest are otherwise regularly deleted. For the original detection of \mtp, we thus had access to a filterbank file with our native resolution of approximately $306.24~\mu$s and 1024 channels across the 856\,MHz bandwidth at L-band. A detailed visual inspection of second-stage candidate plots around the time of discovery showed that we had actually made more detections of the source, but the unusual nature of the wide pulses and the relatively low DM meant that they were not initially labelled as astrophysical and the associated filterbank data had already been deleted. However, there is still sufficient information in the second-stage candidate files to allow the manual determination of an arrival time, albeit with significantly increased uncertainty; we thus obtained a total of 14 approximate arrival times from the day of the discovery.

The uncertainties on those times of arrival (ToAs) were estimated as part of the initial periodicity analysis, which consisted of a simple Bayesian linear regression with three parameters: a pulse period, an initial phase term, and an uncertainty scaling factor where the underlying assumption was that the uncertainty on each arrival time was proportional to the pulse width reported by the search pipeline. This yielded the initial period estimate of $P = 75.89 \pm 0.01$ seconds previously mentioned, and a root mean square uncertainty of 100 ms on the arrival times.
These 14 initial ToAs were then used along with the imaging data to further constrain the pulse period. They are also included in the timing analysis (see the orange points in Extended Data Figure \ref{fig:timing}). 

We generated ToAs for each of the two 30 minutes observations on each of the 6 MeerKAT L-band observing epochs between February and May 2021. The filterbank data recorded with the TUSE instrument were used for timing. These data also had a resolution of approximately $306.24\mu$s and 1024 channels across the 856\,MHz bandwidth at L-band. The data were folded on to 65536 phase bins and incoherently dedispersed using the \textsc{dspsr} software package \citep{2011VanStraten}. Further manipulation of the data used the tools available in the \textsc{psrchive} package \citep{Hotan+2004}. Radio-frequency interference was removed manually using \textsc{psrzap}.

To maintain a coherent timing solution, we secured available observing slots at the Parkes radio telescope in Australia and monitored \mtp\, over 4 epochs. The observations were performed using the Ultra-Wideband receiver (UWL) across a 704--4032~MHz band. The data were coherently de-dispersed at the DM given in Table \ref{tab:timingparams}, divided into 1024 phase bins for each of 3328 frequency channels of 1~MHz bandwidth, and written to disk. The duration of the observations varied between 1 and 2 hours. 
The RFI environment at Parkes is such that many frequency channels in the data contained strong baseline fluctuations on a time scale comparable to the typical pulse duration of \mtp, which required applying additional layers of RFI mitigation to make the averaged pulse unambiguously identifiable in each integrated observation. An updated version of the \texttt{clfd} (\url{https://github.com/v-morello/clfd}) RFI cleaning package described in \citep{MBC+19} was used to that effect. The relatively steep spectrum of the source meant that it was not easily detectable above 1.8~GHz even after cleaning and so we only used frequencies in the range 0.704 to 1.8~GHz GHz for generating arrival times. 

The initial folding and timing analysis for both telescopes used the best known period and DM at the time and the position determined from the imaging. For the MeerKAT data a noise-free template was made by fitting von Mises functions to the data from the first epoch in February 2021 using the \textsc{psrchive} program \textsc{paas}. A similar procedure was followed for the Parkes UWL data also using the first observation from February 2021. For both the MeerKAT and Parkes data ToAs were obtained by using \textsc{psrchive}'s \textsc{pat} which cross-correlated the template with an average profile for each of the 30 minute observations. A ToA (the first red diamond in Extended Data Figure \ref{fig:timing}) was also determined from the single pulse obtained from the filterbank file saved from the discovery observation by cross-correlating it with the MeerKAT template (As the filterbank file was significantly shorter in duration than the period of the source we padded it with random Gaussian noise so that it could be folded using \textsc{dspsr} and preserve the timing information). 

Timing was done using \textsc{tempo2} \citep{tempo2} with the JPL DE436 planetary ephemeris (\url{https://naif.jpl.nasa.gov/pub/naif/JUNO/kernels/spk/de435s.bsp.lbl}). The ToAs were fit using a model including the period $P$ and period derivative $\dot{P}$ and a jump between the UWL data and the MeerKAT data (Another, fixed jump of 6.115060037383178\,s, was needed for a few of the arrival times from the discovery epoch due to an uncertainty of one data buffer that occurred in these early data). We do not fit for position as it is well determined from the imaging as described in previous sections. An astrometric precision of about $\sim 1$\arcsec\, results in a change in arrival time over a year of less than 2.3~ms, so is not affecting the measured parameters, especially the $\dot{P}$ which can show a covariance when there is less than a year of data. We also do not fit for DM as this is sufficiently well determined from optimizing the S/N of the individual pulses and a jump is fit between the UWL and MeerKAT data. 

Once a coherent timing solution was obtained across the entire data set the filterbanks were refolded and dedispersed and a new noise-free template was made, based on the sum of all of the detected pulses, and new ToAs were obtained and a new final timing solution was determined and is presented in Table \ref{tab:timingparams} with the residuals shown in Extended Data Figure \ref{fig:timing}. Each 30-minute observation is the average of about 24 pulses and so there is some pulse phase jitter, which can be seen in the MeerKAT data, where the error bars are smaller than the data points and also less than the scatter in the arrival times. However the overall timing RMS is only 5\,ms which is just under 1/10000th of the pulse period which is approaching that of the best millisecond pulsars and is attributable to the very high signal to noise but also suggests that the arrival times are not significantly affected by the pronounced variations in the pulse properties and likely reflects the overall similarity of the pulse envelope.

\section*{Quasi-periodicity Analysis}

Radio loud neutron stars are seen to exhibit a rich variety of intensity variations over timescales of microseconds to years. Within an individual pulse, substructures manifest most conspicuously as sub-pulses/components that have random-like but also, pulse-
and source-dependent continuous or quasi-periodic variability in time intervals. Quasi-periodicities
are usually seen as repeating ``micropulses'' \citep{bor76}
forming part of ``microstructure'' superimposed on the wider sub-pulses (e.g.~\citep{lgs12}).
Microstructure is usually theorized to be caused by mechanisms related to magnetospheric radio emission.
There is often a variety of timescales observed, even within a given source (e.g.~\citep{Lange1998}), for pulsars with typical periods of about $\sim 1$ s, 
sub-pulses tend to have widths of a few to tens of ms. Timescales and periodicities 
in the shorter micropulses tend to scale like $ P_\mu \sim P/1000$ \citep{cor79,kjv02}, where often
the micropulse duration and periodicity scale like $w_{\mu} \sim P_\mu / 2$ \cite{cor79}.

We followed standard methods \citep{cor79,Lange1998} to determine the timescales of the
short-time structure for \mtp.
An auto-correlation function (ACF) of pulses detected by the TUSE pipeline containing both sub-pulses and microstructure, will show a peak at zero-lag corresponding to the DC component, followed by a peak at short timescales due to possible microstructure and then a second peak associated with the sub-pulse structure \citep{CWH1990}. We extracted the timescale of the sub-pulses by performing an ACF analysis of the single pulse intensities.  We compute the cross-correlation of the de-dispersed signal as a function of time with a delayed copy of itself given by,
\begin{equation}
    \mathrm{ACF}(\tau) = \int_{0}^{t} f(t) f(t-\tau) \, \mathrm{dt} ,
\end{equation}
where $\tau$ is the time lag. The zero-lag value, associated with self-noise, was excised from the ACF. Given the complexity of the structures seen in the single pules, we visually inspected each pulse to understand what can and cannot be inferred about the timescales. The characteristic separation or quasi-period of the sub-pulses (defined as $P_{2}$) measured across the whole observing band is given by the time lag of the peak of the first feature following the zero-lag in each ACF. We only measure the the quasi-periods for those pulses which are visually obvious in the ACFs. 

As a result, we observe some of the quasi-period values to be harmonically related as shown in Extended Data Figures \ref{fig:acf} and \ref{fig:QP}, and in some cases, the separations between the peaks is almost the separation between the dips or dropouts in power. Occasionally, two or more quasi-periods co-exist in a single rotation. Upon visual inspection we notice that some pulses in \mtp\, exhibit variations in widths as well as quasi-periods within a single rotation as seen in Extended Data Figure \ref{fig:multiperiodACF}. We do not observe these quasi-periodic pulses to follow a trend in time, nor do they appear to precede or follow any other particular type of pulse.  

These quasi-periodic features could be interpreted as sub-pulses or drifting sub-pulses. Although the latter are usually characterized by a fixed, from pulse to pulse, separation between sub-pulses. The sparking discharge from IVG models explains sub-pulses through non-stationary spark-associated plasma flows. The IVG is discharged in the form of dense isolated sparks (i.e. pair-production sites), where the lateral size and distance between the sparks are comparable to the gap height \citep{RS75}. Each spark has a corresponding coherent plasma column, which radiates and generates the observed sub-pulse components in pulsars. According to this `Spark' model \citep{MBM+20}, up to three sparks can be accommodated in the polar cap of \mtp\,, which is much fewer than the number of features typically seen in its single pulses.

The dominant quasi-period of $\sim 76$ ms (corresponding to a frequency of 13 Hz)
follows the spin-period scaling seen in corresponding values
of the micropulses in normal pulsars \citep{cor79} (see Extended Data Figure \ref{fig:QP}). This scaling can be most easily associated with the emission of beamlets making up the wider
sub-pulses \citep{kjv02}, implying that the periodicities are caused by 
a {\em temporal} or {\em angular} mechanism
rather than a {\em radial} mechanism. Apart from the consistent scaling
in the value of the quasi-period, the appearance of several periodicities within one
sub-pulse has also been seen in normal pulsars (e.g.~\citep{bor76,cor79,Lange1998}).
Overall, this makes it tempting to associate these structures to ``normal'' microstructure
in pulsars in a similar manner. 

However, the appearance of the dropouts (see e.g.~the top-left example in Extended Data Figure \ref{fig:acf}) is different
to that of normal micropulses. In contrast, it is very
reminiscent of quasi-periodic oscillation (QPO) features seen both in the emission 
of hard short X-ray bursts and the tail of energetic giant flares of magnetars. The ``dropout'' pattern seen in the quasi-periodic and partially nulling pulses is very unusual for pulsar radio emission. Nevertheless, 
we establish that these dropouts are a genuine feature of the emission of the source. We see these features at all epochs in the filterbank data recorded by both the TUSE and APSUE instruments, as well as the raw voltage data extracted directly from the F-engine. These features track the dispersion of the source and, importantly, are not seen in the ``off pulse" emission. Additionally, we do not observe these features in any of the test pulsar (J0820$-$4114) observations at the start of each epoch, thereby ruling out the possibility of instrumental artifacts. 
The magnetar QPOs with frequencies between 18 to 1800 Hz are 
often interpreted as a result of seismic vibrations of the neutron star (see e.g. \citep{wac+16,kb17} and
references therein). Global magnetoelastic axial (torsional) oscillations are expected to be able
to explain the frequencies as low as seen here. They have also been put forward as explanations for
sequences of emission features seen in Fast Radio Bursts (FRBs) invoking magnetar oscillations 
\citep{WC20}. Adopting such an explanation for the quasi-periodicities observed in \mtp\ would demand,
given their persistence in our observations, a repeating triggering mechanism or modes
with long-lived eigenfrequencies. The nature of the modes, their eigenmodes and, importantly,
their damping times depend strongly on
the physics and properties of the neutron star's crust, the mass, the equation-of-state and, to some
degree, also on magnetic field strength \citep{kb17}. Most references discuss damping time lengths only
for the duration of the seen QPOs or FRB emission sequences, i.e.~tens of ms to seconds (e.g.~\citep{WC20}),
which is clearly too short to explain the consistency of our observed periodicities over
many weeks and months (see Extended Data Figure \ref{fig:QP}).

Interestingly, radio observations of the magnetar XTE~J1810-197 during its renewed radio brightness following its 2018 outburst also revealed a persistent 50-ms periodicity in its pulse profile seen for about 10 days. As reported by Levin et al. (2019)~\citep{lld+19}, this emission feature was imprinted on the pulse
profile simultaneously at a range of observed radio frequencies between 1 and 9 GHz. The frequency
of this feature of about 20 Hz is obviously similar to that of our dropouts in \mtp. In
XTE~J1810$-$197 the feature showed a remarkable constancy in phase relative to the main pulse profile,
implying that the periodicity is not a temporal modulation of the emitting source, but must be due to a periodic structure in the radiation beam pattern that sweeps across the Earth as the pulsar rotates.
Levin et al.~suggested that
this pattern could arise from a {\em stable} structure on the surface of the neutron star at the base of the magnetic field lines hosting the emitting particles for the radio component. 
The stability of the pattern would require a frozen-in wave pattern of radial dimension
of the surface height, temperature or magnetic field. Such a pattern would be reminiscent of surface waves in the neutron star crust, similar to those discussed above, but they would need to be stable over at least
10 days.

\section*{X-ray Follow-up}
% \td{SM}

We requested Neil Gehrels \textit{Swift} Observatory (\textit{Swift}) observations to search for a candidate X-ray counterpart to \mtp. We obtained three observations for a total exposure of 7419.030~s (ObsIDs 00014019002, 00014019003, 00014019004, taken on MJD 59245.83, 59352.04, and 59353.29 with individual exposures of 3872.021~s, 1345.595~s, and 2703.971~s, respectively). The three observations were separated by $1$, $9$, and $10$ days from their closest radio observation, respectively. We extracted an image using the XRT product generator online reduction pipeline (\url{https://www.swift.ac.uk/user\_objects/index.php}) \citep{Evans2009}. A visual inspection of the image showed that the source was not detected, therefore we use the \textsc{SOSTA} tool within the \textsc{XIMAGE} environment to perform source statistics. \textsc{SOSTA} allows one to use a local background to determine the significance of a source and its count rate, rather than a global background estimate. Extracting events at the nominal position of \mtp\, from a square box with 16 pixels side ($\approx$38\arcsec) we obtained a 3-sigma upper limit to the count rate of R$^{up}$ $\approx$ 1.57$\times 10^{-3}$ cts/s in the 0.5$-$10 keV energy range.
We then assumed a blackbody spectrum with temperature 1.5 keV, and an equivalent column density of  N$_{\rm H}$ = 4.32$\times$10$^{21}$ cm$^{-3}$ (i.e. the Galactic equivalent column density in the direction of \mtp). Using WebPPIMS (\url{https://heasarc.gsfc.nasa.gov/cgi-bin/Tools/w3pimms/w3pimms.pl}, PIMMS v4.11b.) we estimated a 3-sigma upper limit to the 0.5$-$10 keV X-ray flux of F $<$ 1.2$\times$10$^{-13}$ erg cm$^{-2}$ s$^{-1}$, which at a distance of d$_1$ $\approx$ 328~pc and d$_2$ $\approx$ 467~pc corresponds to a 3-sigma upper limit to the X-ray luminosity L$_{\rm X1}$ $<$ 1.6$\times$10$^{30}$ erg s$^{-1}$ and L$_{\rm X2}$ $<$ 3.2$\times$10$^{30}$ erg s$^{-1}$, respectively. 

We note that the XMM-Newton archive includes two archival observations of the Vela X-1 field (ObsID 0406430201 and 0841890201, taken in 2018 and 2019, respectively). In such pointings \mtp\ is located at the very edge of the EPIC-MOS image, and in one of the two observations only one of the MOS cameras was active (the other was switched off for telemetry reasons). Given that the response of the instrument is not ideal so close to the edge of the CCD, and calibration might not be reliable, we decided to not use these observations, and rely solely on the more conservative - but likely more robust - upper limit derived from the \textit{Swift} data.

%%%%%%%%%%%%%%==============================%%%%%%%%%%%%%%
%%%%%%%%%%%%%%==============================%%%%%%%%%%%%%%
%%%%%%%%%%%%%%==============================%%%%%%%%%%%%%%

\backmatter

\bmhead{Acknowledgments}

This manuscript makes use of MeerKAT (Project ID: DDT-20210125-MC-01) and Parkes data (Project ID: PX071). M.C. would like to thank SARAO for the approval of the MeerKAT DDT request, and the science operations, CAM/CBF and operator teams for their time and effort invested in the observations. The MeerKAT telescope is operated by the South African Radio Astronomy Observatory, which is a facility of the National Research Foundation, an agency of the Department of Science and Innovation (DSI). The Parkes Radio Telescope (\textit{Murriyang}) is managed by CSIRO. We acknowledge the Wiradjuri people as the traditional owners of the Parkes observatory site. M.C. would like to thank the ATNF for scheduling observations with the Parkes radio telescope. The SALT observations were obtained under the SALT Large Science Programme on transients (2018-2-LSP-001; PI: DAHB) which is also supported by Poland under grant no. MNiSW DIR/WK/2016/07. 

\section*{Declarations}

% Some journals require declarations to be submitted in a standardised format. Please check the Instructions for Authors of the journal to which you are submitting to see if you need to complete this section. If yes, your manuscript must contain the following sections under the heading `Declarations':

\begin{itemize}

\item \textbf{Funding}
M.C., B.W.S., K.R., M.M., V.M., S.S., F.J., M.S., L.N.D, and M.C.B. acknowledge funding from the European Research Council (ERC) under the European Union's Horizon 2020 research and innovation programme (grant agreement No 694745). M.C. acknowledges support of an Australian Research Council Discovery Early Career Research Award (project number DE220100819) funded by the Australian Government and the Australian Research Council Centre of Excellence for All Sky Astrophysics in 3 Dimensions (ASTRO 3D), through project number CE170100013. K.R. acknowledges support from the Vici research program `ARGO' with project number 639.043.815, financed by the Dutch Research Council (NWO). J.v.d.E. is supported by a Lee Hysan Junior Research Fellowship awarded by St. Hilda’s College, Oxford. DAHB and P. Woudt acknowledge research support from the National Research Foundation. 

\item \textbf{Data availability} The data that support the findings of this study are available at \url{https://github.com/manishacaleb/MKT-J0901-4046}.

\item \textbf{Code availability} All code necessary for analyses of the data are available on GitHub and Zenodo: \url{https://github.com/IanHeywood/oxkat}, \url{https://doi.org/10.5281/zenodo.1212487}

\item \textbf{Author contributions}
M.C. and B.W.S drafted the manuscript with suggestions from co-authors. M.C. is PI of the MeerKAT DDT and Parkes data. B.W.S. is PI of MeerTRAP and R.F. and P.W. are PIs of the ThunderKAT data. M.C. reduced and analyzed the radio time domain data for quasi-periodicity, and M.C. and M.K. interpreted it. I.H. calibrated, imaged and performed astrometry on the data to localize the source. B.W.S., V.M. and F.J. undertook the timing analyses. E.B. and K.R. designed and built the complex channelized data capture system. K.R. and P. Weltevrede performed the polarization analyses. M.M. carried out the pulse-width analyses using the wavelet transform method. E.B. and W.C. built and designed the beamformer used by MeerTRAP. J.v.d.E. and S.M. performed the Swift analysis. D.A.H.B., J.B. and P.W. obtained and analyzed data from the SALT and SAAO-1m telescopes. D.A.H.B. acknowledges support from the Natioanal Research Foundation. F.J. and M.S. undertook analysis of the extant data. S.B. assisted in planning and scheduling the MeerKAT observations. S.S., F.J., M.S., R.F., L.N.D. and M.C.B contributed to discussions about the nature of the source.

\item \textbf{Conflict of interest/Competing interests} 
The authors declare no competing interests.

\end{itemize}

\clearpage

%%%%%%%%%%%%%%%%%%%%%%%%%%%%%%%%%%%%%%%%%%%%%%%%%%%%%%%%%%%%%%%%%%%%%%%
% -------- Beginning of main text and extended data tables % -----------------
%%%%%%%%%%%%%%%%%%%%%%%%%%%%%%%%%%%%%%%%%%%%%%%%%%%%%%%%%%%%%%%%%%%%%%%

%00000000000000000000000%
% Main text Table 1
%00000000000000000000000%

\begin{center}
\begin{table}
\caption{Pulsar timing and model parameters for \mtp\,. This includes the measured quantities and the derived quantities from the timing analysis over the span of this observing campaign. Uncertainties in parentheses as 1-$\sigma$ errors
on the last significant quoted digit.}
\begin{tabular}{ll}
\hline\hline
\multicolumn{2}{c}{Data and model fit quality} \\
\hline
Modified Julian Date (MJD) range\dotfill & 59119.0 to 59343.6 (7.4 months) \\ 
Number of TOAs\dotfill & 29 \\
Weighted root mean square timing residual (ms)\dotfill & 5.7 \\

\hline

\multicolumn{2}{c}{Measured quantities} \\ 
\hline
Right ascension, $\alpha$ (J2000)\dotfill &  $09^h01^m29.249^s \pm 1.0''$ \\ 
Declination, $\delta$ (J2000)\dotfill & $-40^{\circ}46'02.984'' \pm 1.0''$ \\ 
Pulse frequency, $\nu$ \dotfill & $0.013177739873 \pm 9.9 \times 10^{-12}$~s$^{-1}$ \\ 
First derivative of pulse frequency, $\dot{\nu}$ \dotfill & $-3.9 \pm 0.2$~s$^{-2}$ \\ 
Pulse period, $P$ \dotfill & 75.88554711 $\pm \, (6 \times 10^{-8})$~s\\
Period derivative, $\dot{P}$ \dotfill & ($2.25 \pm 0.1) \times 10^{-13}$ s s$^{-1}$ \\
Dispersion measure, DM \dotfill & $52 \pm 1 \, \rm{pc \, cm^{-3}}$\\
Full width at half maximum, $W_{50}$ (L-band) \dotfill & $299 \pm 1$~ms\\ 
Full width at half maximum, $W_{50}$ (UHF-band) \dotfill & $296 \pm 4$~ms\\ 
Spectral index, $\alpha$ \dotfill & $-1.7 \pm0.9$ \\ 
Rotation measure, (RM) \dotfill & $-64 \pm 2$ rad~m$^{-2}$\\ 
Fractional linear polarization \dotfill & $12.2 \pm 0.2$~\% \\
Fractional circular polarization \dotfill & $21.0 \pm 1.9$~\% \\
\hline
\multicolumn{2}{c}{Inferred quantities} \\ 
\hline
Distance (\textsc{ymw16}), $d_{1}$ \dotfill & 328~pc\\
Distance (\textsc{ne2001}), $d_{2}$ \dotfill & 467~pc\\
Characteristic age, $\tau$ \dotfill & 5.3~Myr\\
Surface dipole magnetic field strength, $B$  \dotfill & $1.3\times 10^{14}$~G\\
Spin-down luminosity, $\dot{E}$ \dotfill & $2.0\times10^{28}$~erg~$\mathrm{s}^{-1}$ \\
Period-averaged radio luminosity, L$_{\rm 1400}$ at $d_{2}$ \dotfill & $ 89\, \mu\mathrm{Jy\,kpc^{2}}$ \\
X-ray Luminosity, L$_{\rm X} (0.5-10 \, \rm {keV})$ at $d_{2}$ \dotfill &  $\lesssim 3.2\times$10$^{30}$ erg s$^{-1}$ \\

\hline
\end{tabular}
\label{tab:timingparams}
\end{table}
\end{center}

%00000000000000000000000%
% Main text Table 2
%00000000000000000000000%

% \setcounter{table}{0}
% \captionsetup[table]{name={\bf Extended Data Table}}

\begin{table*}
\footnotesize
\centering
\caption{MeerKAT observations of the \mtp\, field. The first three rows labelled TKAT are discovery observations targeting the Vela X-1 field, while the rest labelled DDT are follow-up observations. See the text for details.}
\begin{tabular}{lllllllll} 
\hline
Date        & Block ID   & RA                          & Dec                                & Band & N$_{\mathrm{ant}}$ & T$_{\mathrm{obs}}$  & T$_{\mathrm{int}}$ & Origin     \\ 
UT, J2000   &            & J2000                       & J2000                              &      &     & h                   & s    &            \\ \hline
2020-09-25  & 1600995961 & 09$^{h}$02$^{m}$06.86$^{s}$ & $-$40$^{\circ}$33$^{'}$16.9$^{''}$ & L    & 59    & 0.5                 & 8    & TKAT \\
2020-09-27  & 1601168939 & 09$^{h}$02$^{m}$06.86$^{s}$ & $-$40$^{\circ}$33$^{'}$16.9$^{''}$ & L    & 61    & 0.5                 & 8    & TKAT \\
2020-10-11  & 1602387062 & 09$^{h}$02$^{m}$06.86$^{s}$ & $-$40$^{\circ}$33$^{'}$16.9$^{''}$ & L    & 60    & 0.5                 & 8    & TKAT \\ 
2021-02-01  & 1612141271 & 09$^{h}$01$^{m}$29.35$^{s}$ & $-$40$^{\circ}$46$^{'}$03.6$^{''}$ & L    & 64    & 1                   & 2    & DDT \\
2021-02-02  & 1612227667 & 09$^{h}$01$^{m}$29.35$^{s}$ & $-$40$^{\circ}$46$^{'}$03.6$^{''}$ & L    & 61    & 1                   & 2    & DDT \\
2021-02-10  & 1612994791 & 09$^{h}$01$^{m}$29.35$^{s}$ & $-$40$^{\circ}$46$^{'}$03.6$^{''}$ & L    & 62    & 1                   & 2    & DDT \\
2021-03-03  & 1614794470 & 09$^{h}$01$^{m}$29.35$^{s}$ & $-$40$^{\circ}$46$^{'}$03.6$^{''}$ & L    & 63    & 1                   & 2    & DDT \\
2021-04-02  & 1617367872 & 09$^{h}$01$^{m}$29.35$^{s}$ & $-$40$^{\circ}$46$^{'}$03.6$^{''}$ & L    & 63    & 1                   & 2    & DDT \\
2021-04-02  & 1617376889 & 09$^{h}$01$^{m}$29.35$^{s}$ & $-$40$^{\circ}$46$^{'}$03.6$^{''}$ & UHF  & 62    & 1                   & 2    & DDT \\ 
2021-05-09  & 1620567645 & 09$^{h}$01$^{m}$29.35$^{s}$ & $-$40$^{\circ}$46$^{'}$03.6$^{''}$ & L    & 62    & 1                   & 2    & DDT \\ \hline
\end{tabular}
\label{tab:observations}
\end{table*}

%%%%%%%%%%%%%%%%%%%%%%%%%%%%%%%%%%%%%%%%%%%%%%%%%%%%%%%%%%%%%%%%%%%%%%%%%%
% -------- End of main text and extended data tables % -------------------
%%%%%%%%%%%%%%%%%%%%%%%%%%%%%%%%%%%%%%%%%%%%%%%%%%%%%%%%%%%%%%%%%%%%%%%%%%

\clearpage

%%%%%%%%%%%%%%%%%%%%%%%%%%%%%%%%%%%%%%%%%%%%%%%%%%%%%%%%%%%%%%%%%%%%%%%%%%
% -------- Beginning of main text and extended data figures % ------------
%%%%%%%%%%%%%%%%%%%%%%%%%%%%%%%%%%%%%%%%%%%%%%%%%%%%%%%%%%%%%%%%%%%%%%%%%%

%00000000000000000000000%
% Main text Figure 1
%00000000000000000000000%

%00000000000000000000000%
% Main text Figure 2
%00000000000000000000000%

%00000000000000000000000%
% Extended data Figure 1
%00000000000000000000000%

\setcounter{figure}{0}
\captionsetup[figure]{name={\bf Extended Data Figure}}

\begin{figure}
    \centering
    \includegraphics[width=4.0 in]{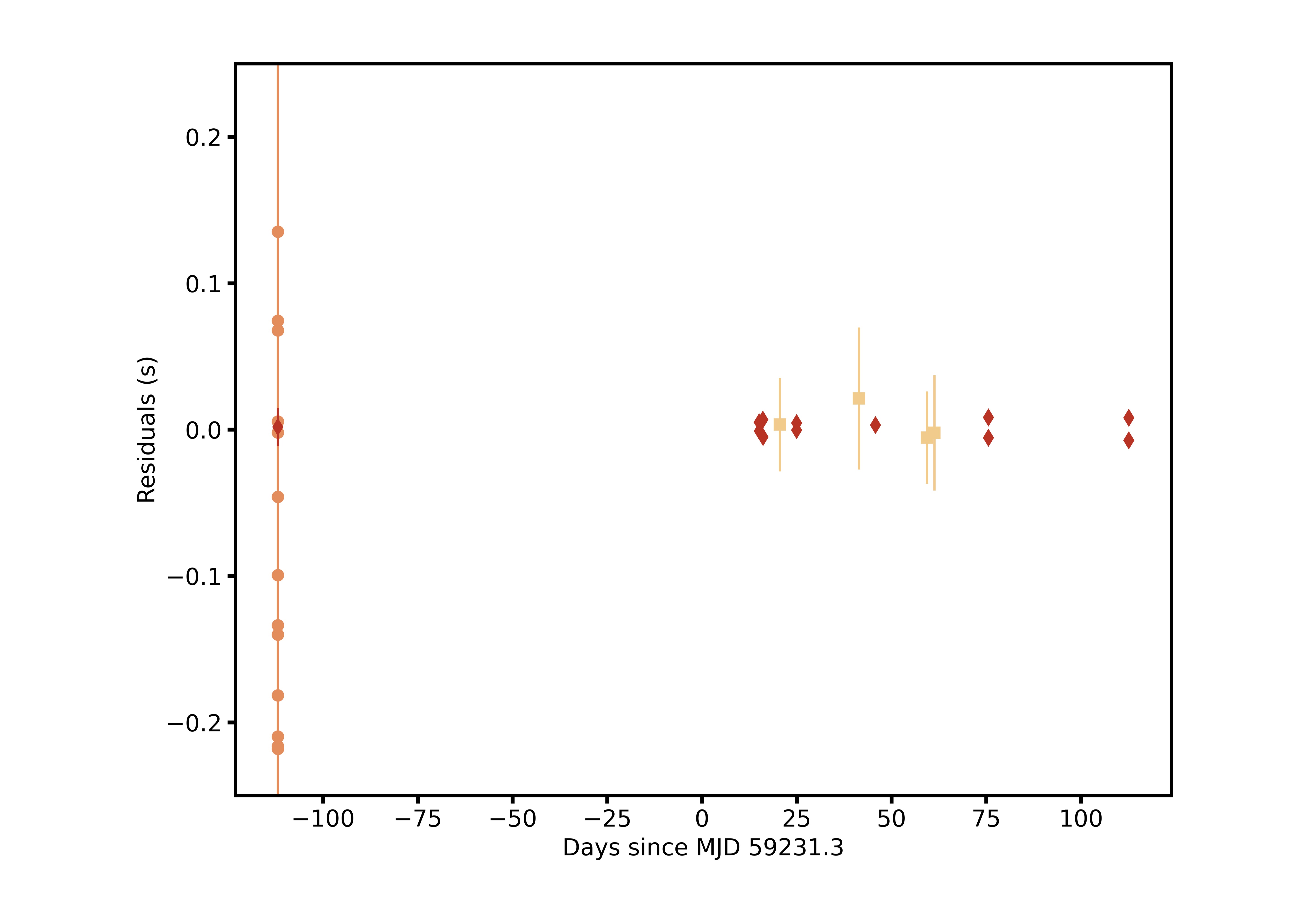}
    \caption{Timing residuals of \mtp. The residuals from the best fit timing model given in Table \ref{tab:timingparams}. The orange data points are determined from the original MeerTRAP detection images, the first red diamond corresponds to a single pulse and the remaining red diamonds are determined from each of the half hour long follow-up observations with MeerKAT. The error bars are 1-$\sigma$. We used the L-band MeerKAT data for the timing analysis. The light coloured data points are from the Parkes UWL observations.}
    \label{fig:timing}
\end{figure}

%00000000000000000000000%
% Extended data Figure 2
%00000000000000000000000%

\begin{figure}
    \centering
    \includegraphics[width=\linewidth]{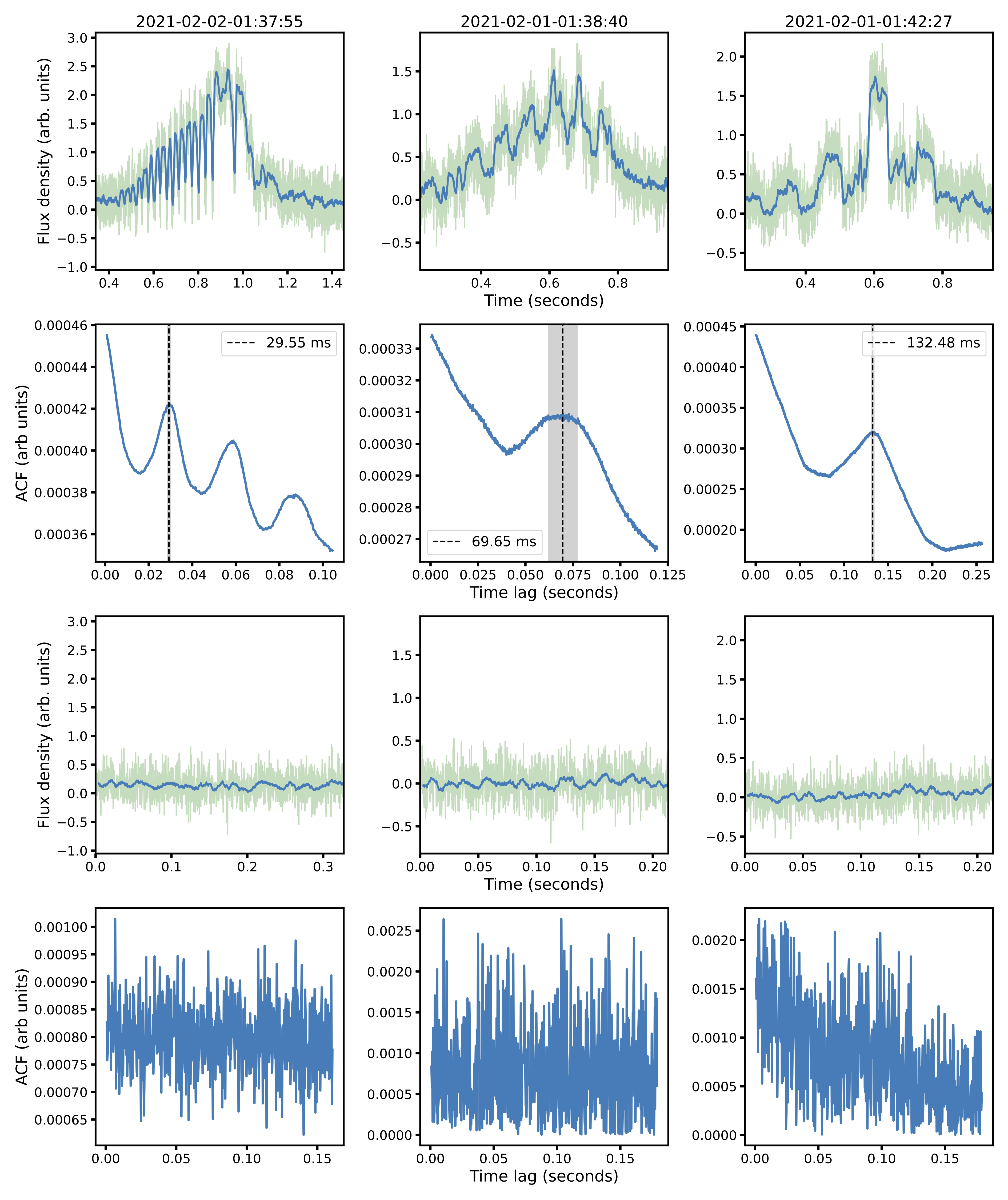}
    \caption{Examples of quasi-periodic pulses. The top two rows show pulse profiles and their corresponding ACFs at 306.24$\mu$s resolution, respectively. The value the of quasi-period is indicated by the black vertical lines. The bottom two rows show the off-pulse regions and their corresponding ACFs.}
    \label{fig:acf}
\end{figure}

%00000000000000000000000%
% Extended data Figure 3
%00000000000000000000000%

% \begin{sidewaysfigure}
\begin{figure}
    \centering
    \includegraphics[width=4in]{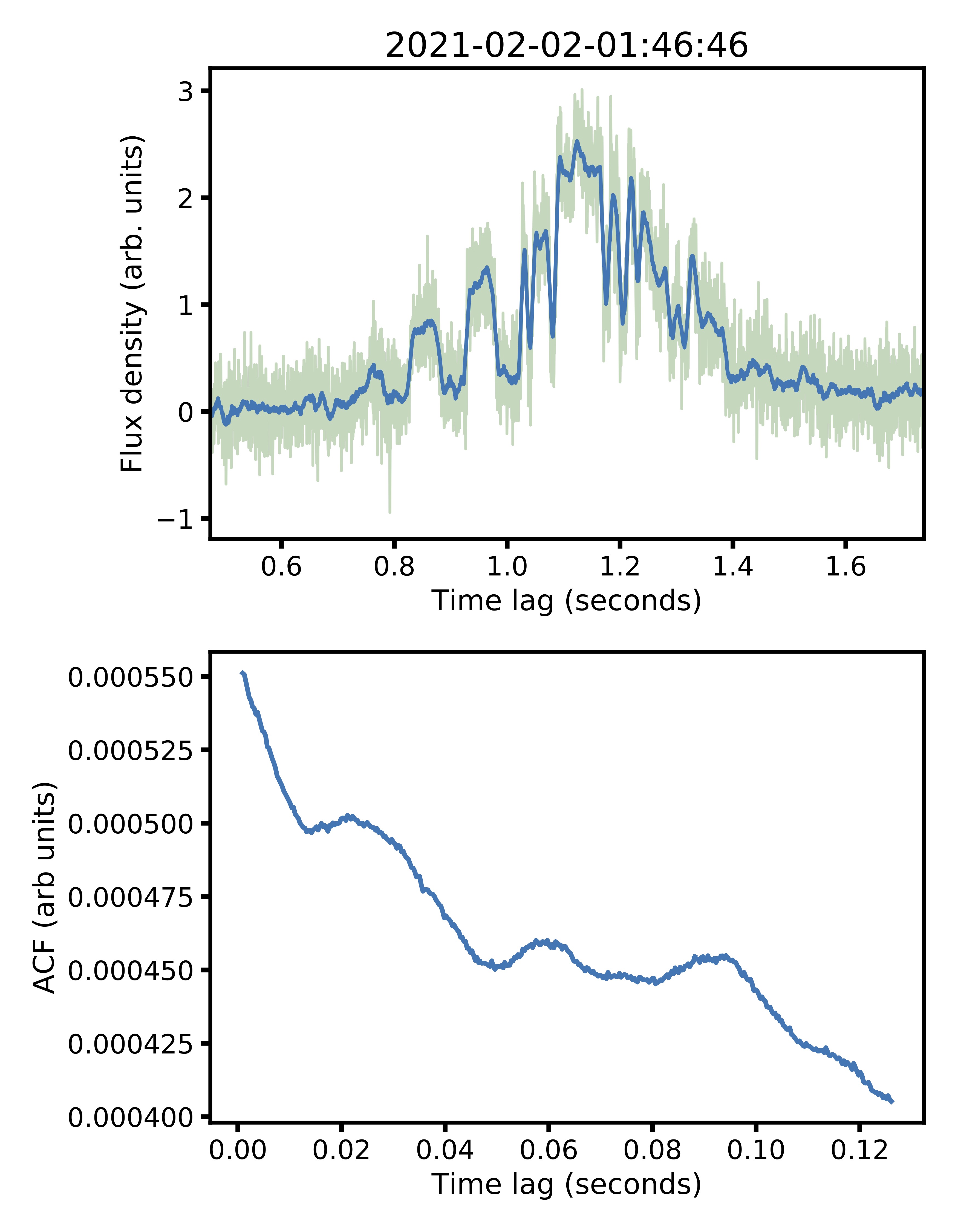}    
    \caption{Example of a pulse exhibiting more than one quasi-period. Some quasi-periodic pulses as shown here, exhibit multiple quasi-periods within a single rotation.}
    \label{fig:multiperiodACF}
\end{figure}

%00000000000000000000000%
% Extended data Figure 4
%00000000000000000000000%

\begin{figure}
    \centering
    \includegraphics[width=4in]{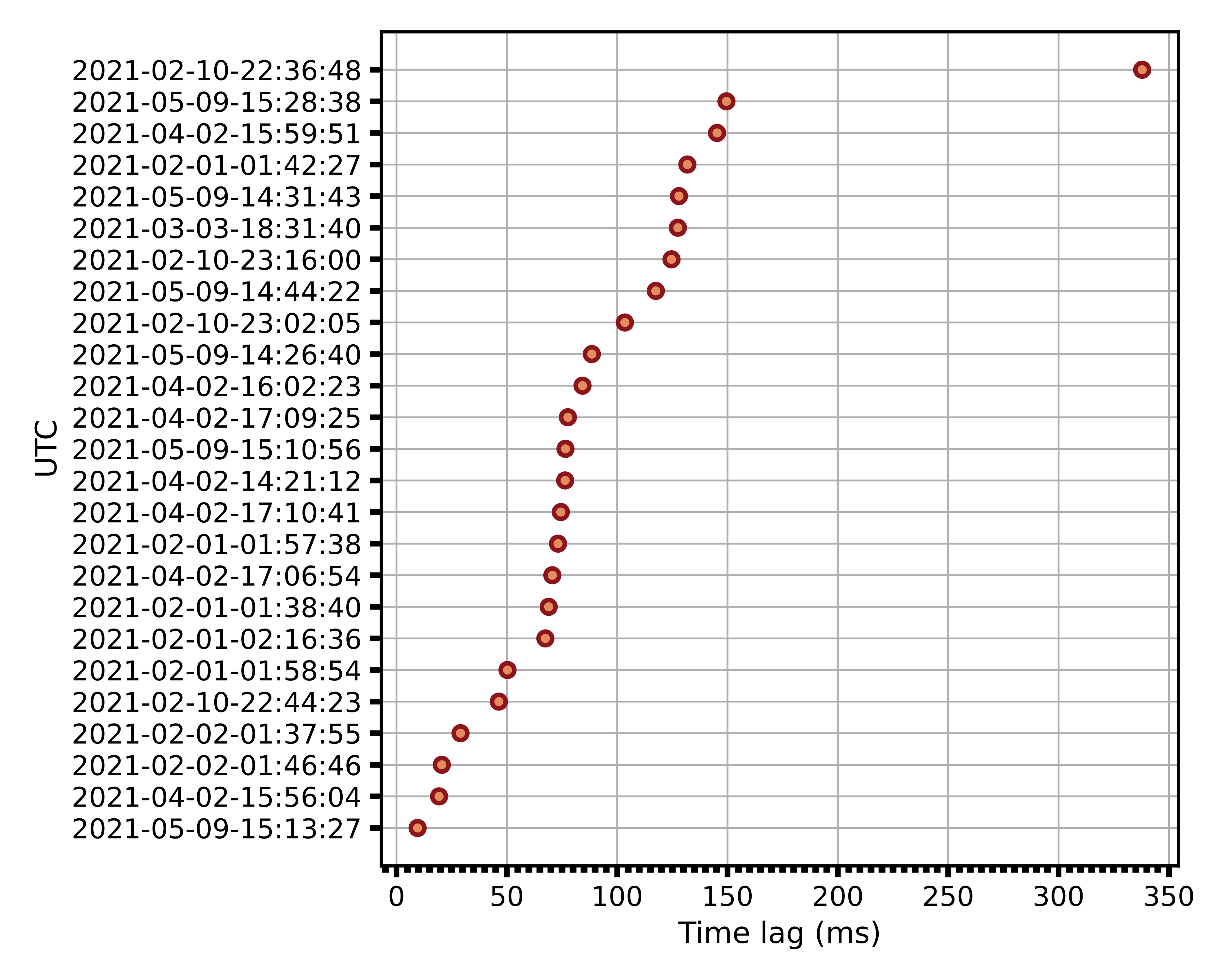}
    \caption{Estimates of the quasi-period across all epochs. The (orange) circles are the measured quasi-periods for each single pulse. The most commonly observed average quasi-period is 75.82~ms with the minimum period being 9.57~ms. The lags are arranged in lag length and not in time order.}
    \label{fig:QP}
\end{figure}

%00000000000000000000000%
% Extended data Figure 5
%00000000000000000000000%

\begin{figure}
    \centering
    \includegraphics[width=11cm]{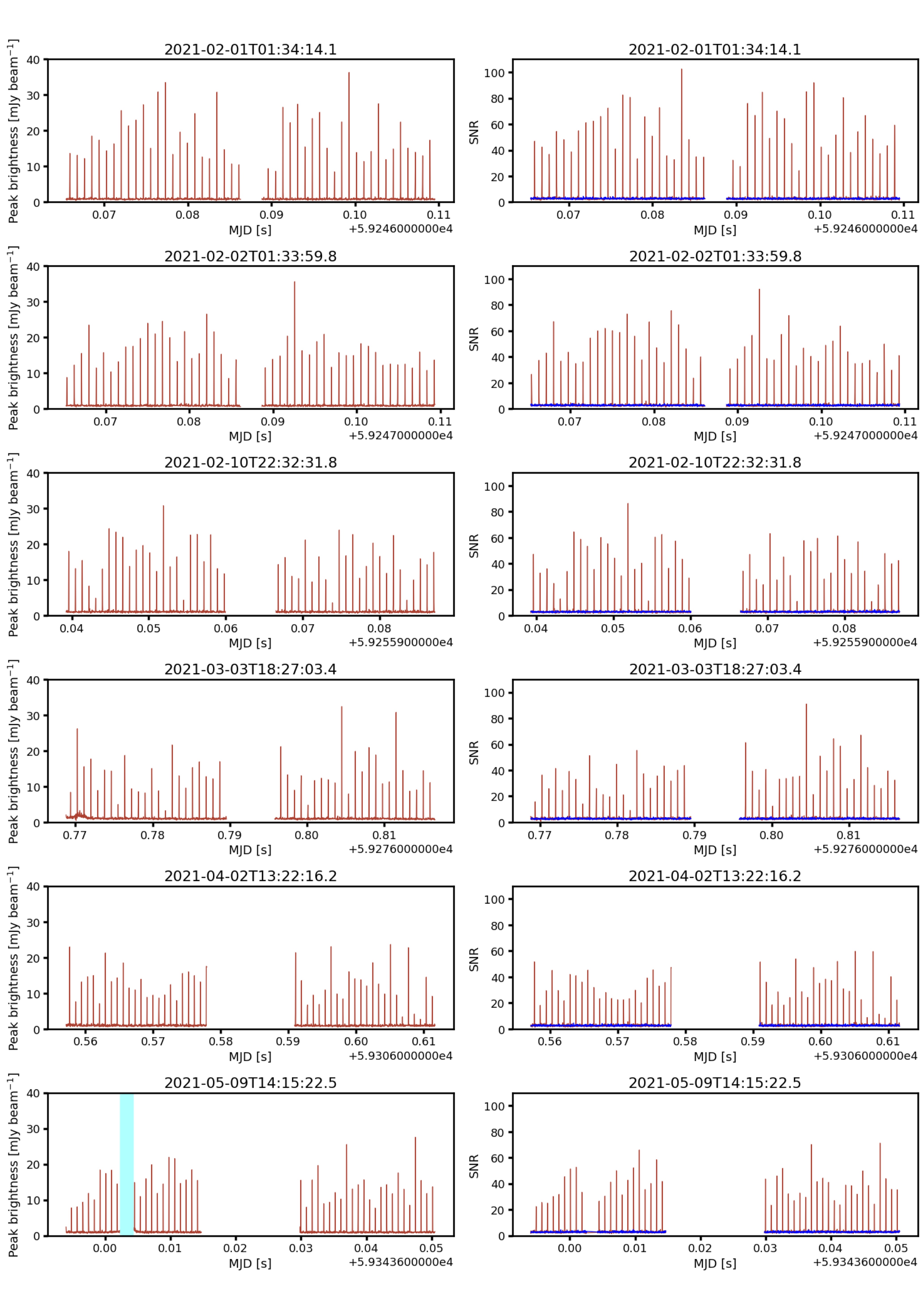}%
    \caption{Radio light-curves of \mtp\,. A regular series of pulsed emission detected in the L-band snapshot imaging for six observing epochs. Please refer to Section \ref{sec:snapshot-imaging} for details.}
    \label{fig:lightcurves}
\end{figure}

%00000000000000000000000%
% Extended data Figure 6
%00000000000000000000000%

\begin{figure}
    \centering
    \includegraphics[width=4.5 in]{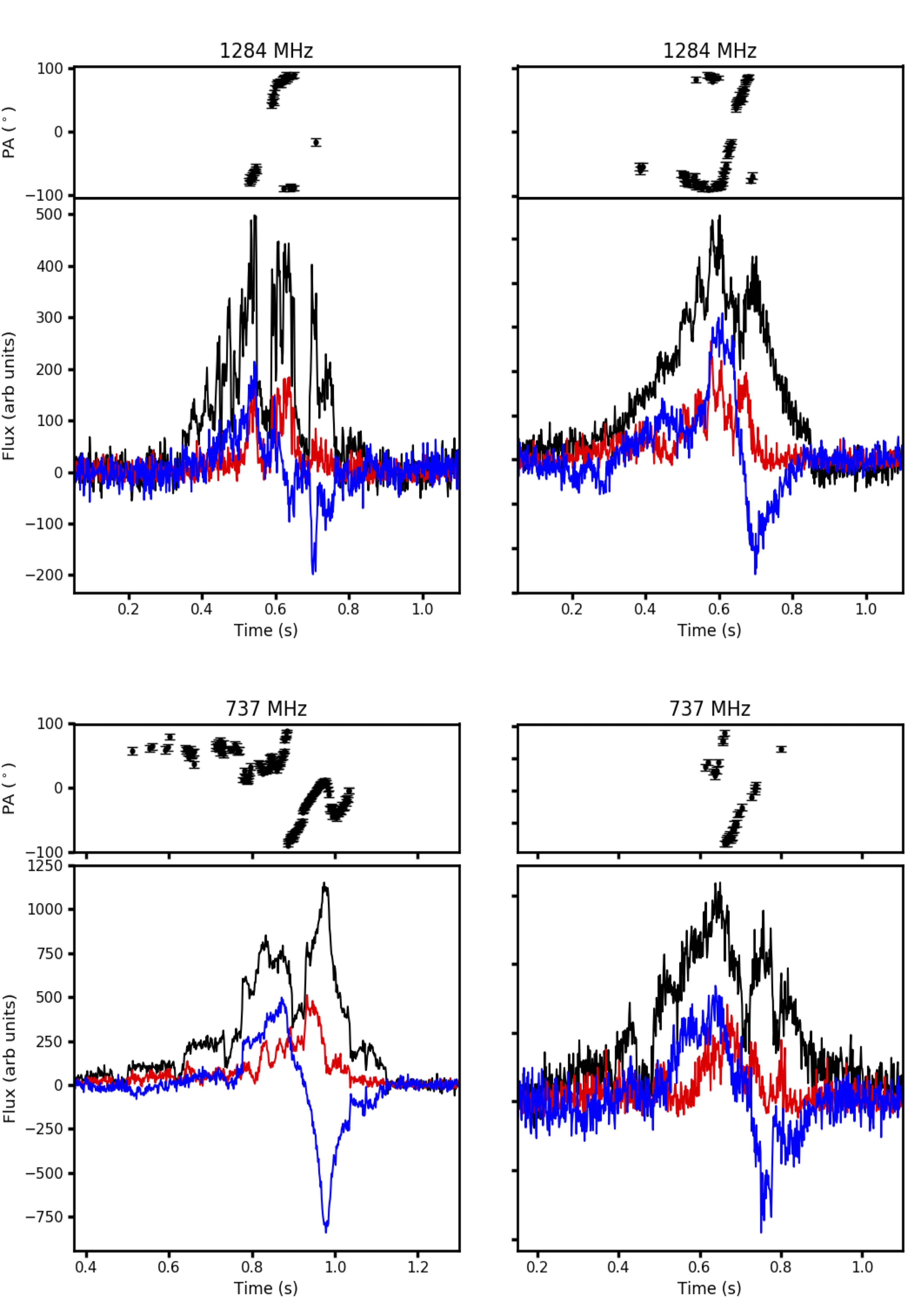}\\
    \caption{Polarization profiles of \mtp\, at 1.3~GHz and 700~MHz. Top Panel: Time series of two single pulses of \mtp\, at 1284 MHz. Bottom Panel: Two different single pulse time series at 737~MHz. For both panels, the total intensity is represented by the black solid line, the red solid line denotes the linear polarization while the blue solid line denotes circular polarization. The polarization position angle is not absolutely calibrated at 737~MHz.}
    \label{fig:polnprof}
\end{figure}

%00000000000000000000000%
% Extended data Figure 7
%00000000000000000000000%

\begin{figure}
    \centering
    \includegraphics[width=5 in]{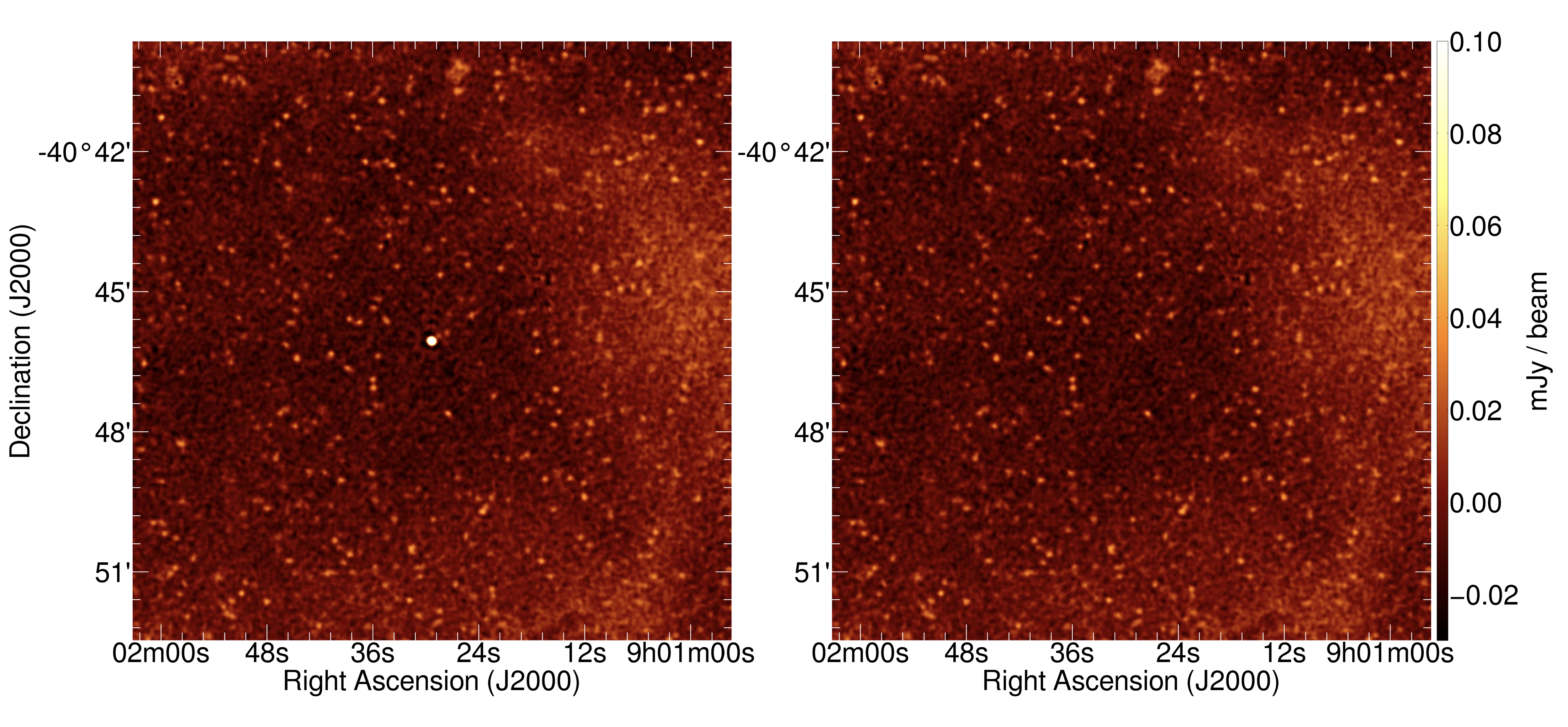}    
    \caption{MeerKAT image of the \mtp\, region at 1.28~GHz. The left hand panel shows the image with the pulsed emission included, and the right hand panel shows the same field following the removal of the integration times containing pulses. No persistent radio source is associated with \mtp\, to a 3$\sigma$ limit of 18 $\mu$Jy beam$^{-1}$. The diffuse shell-like structure that surrounds \mtp\, is partially visible, possibly the supernova remnant from the event that formed the neutron star.}
    \label{fig:OnOff}
\end{figure}

%%%%%%%%%%%%%%%%%%%%%%%%%%%%%%%%%%%%%%%%%%%%%%%%%%%%%%%%%%%%%%%%%%%%%%%
% -------- End of main text and extended data figures -----------------
%%%%%%%%%%%%%%%%%%%%%%%%%%%%%%%%%%%%%%%%%%%%%%%%%%%%%%%%%%%%%%%%%%%%%%%

\clearpage

% \bibliography{bib.bib}% common bib file

%%%%%%%%%%%%%%==============================%%%%%%%%%%%%%%%%%%%%%%%%%%%%==============================%%%%%%%%%%%%%%%%%%%%%%%%%%%%==============================%%%%%%%%%%%%%%

%--------------------- Start of Supplementary Information ---------------------------------%

%%%%%%%%%%%%%%==============================%%%%%%%%%%%%%%%%%%%%%%%%%%%%==============================%%%%%%%%%%%%%%%%%%%%%%%%%%%%==============================%%%%%%%%%%%%%%

\clearpage
\newpage
\section*{Supplementary Information}
\onecolumn

\setcounter{figure}{0}
\captionsetup[figure]{name={\bf Supplementary Figure}}

\section*{MeerKAT description and configuration for transient searches}

The Meer(more) Karoo Array Telescope (MeerKAT) \citep{Jonas, Camilo18, Mauch} operated by the  South African Radio Astronomy Observatory (SARAO) comprises 64, 13.5-m antennas distributed over 8-km in the Karoo region in South Africa. 40 of these dishes are concentrated in the inner $\sim1$-km core. The MeerTRAP project is a commensal programme to search for pulsars and fast transients whilst piggy-backing on the large survey programmes at MeerKAT. The ThunderKAT programme is the image-plane transients search programme at MeerKAT and operates in synergy with MeerTRAP for blind transient searches. In the observations presented in this work, MeerKAT operated at a centre frequency of 1284~MHz with a usable bandwidth of $\sim770$~MHz in the L-band observations and at a centre frequency of 798~MHz with a usable bandwidth of $\sim435$~MHz in the UHF-band observations. MeerKAT observes simultaneously in incoherent and coherent modes using the MeerTRAP backend. In the coherent mode, the voltages from the inner 40 dishes of the $\sim1$-km core of the array are coherently combined to form up to 780 beams on sky with an aggregate field-of-view of $\sim0.4$~deg$^{-2}$. In the incoherent mode the intensities of all available (up to the maximum 64) MeerKAT dishes are added to create a less sensitive but much wider field-of-view of $\sim$1.3~deg$^{2}$. 

The MeerTRAP backend is the association of two systems: the Filterbank and Beamforming User Supplied Equipment (FBFUSE), a many-beam beamformer that was designed and developed at the Max-Planck-Institut f{\"u}r Radioastronomie in Bonn~\citep{barr2018, CBK+21}, and the Transient User Supplied Equipment (TUSE), a real-time transient detection instrument developed by the MeerTRAP team at the University of Manchester \citep{CSA+20}. The input data stream to FBFUSE consists of the complex-valued channels from every dish produced by the MeerKAT F-engine, which are transmitted via the Central BeamFormer (CBF) network; FBFUSE applies the geometric and phase delays (obtained by observing a bright calibrator) before combining the data streams from the dishes into one incoherent beam and up to 780 total intensity tied-array beams. The beams can be placed at any desired locations within the primary beam of the array, but are by default tessellated into a circular tiling centered on the boresight position, and spaced so that the response patterns of neighbouring beams intersect at the 25\% peak power point. The beams are then sent over the network to TUSE for processing.

\section*{Transient search pipeline}
TUSE consists of 67 Lenovo servers with one head node and 66 compute nodes. Each compute node contains two Intel Xeon CPUs  with 16 logical cores each, two Nvidia GeForce 1080 Ti Graphical processing units (GPUs) and 256 GB of DDR4 Random Access Memory (RAM). Each of the nodes is connected to a breakout switch via 10 GbE network interface cards (NIC) that are used to ingest data from FBFUSE. The data from FBFUSE are received on the NICs as  SPEAD2 (\url{https://casper.ssl.berkeley.edu/wiki/SPEAD}) heaps: small, self-describing units that contain a header and 8KB of data in time-frequency order, attached to a specific frequency sub-band and beam. FBFUSE streams these heaps over a range of multicast network addresses, where each address is associated with a fixed subset of 12 coherent beams in a typical configuration. Each TUSE node then subscribes to a distinct multi-cast address, which effectively distributes the processing load evenly across the TUSE cluster. The data ingest code on a TUSE node collects heaps and writes their contents to POSIX shared memory ring buffers (one per beam), so that the data for each beam are reconstructed in their natural time-frequency order (i.e. with the frequency axis contiguous in memory). Once a data segment has been fully reconstructed, it is ingested by the search pipeline. More details on TUSE will be presented in an upcoming paper.

We utilize the highly-optimized Graphics Processing Unit (GPU)-based \textsc{astroaccelerate} software \citep{armour2011gpu, adamek2020single} to search for dispersed signals. The real-time search is performed by incoherently de-dispersing in the DM range 0--5118.4~pc cm$^{-3}$ at L-band and 0--2664.9~pc cm$^{-3}$ in the UHF-band, divided into multiple sub-ranges with progressively increasing DM steps and time averaging factors. We also search up to maximum boxcar widths of 0.67~s and 0.77~s in the L-band and UHF-band respectively. The extracted candidate files contain raw filterbank data of the dispersed pulse and additional padding of 0.5~s at the start and at the end of the file.

For the targeted observations, where the position of \mtp\, is known to within an arcsecond, we were able to run in a mode where only 2 nodes with one beam per node were processing the data in real time. Only extracted candidates were saved for further examination. In addition to the real-time processing, two more nodes were used to record the continuous data to disk at a full data rate for further offline processing.

\section*{APSUSE data recording}
The APSUSE instrument consists of 68 high-performance Huawei FusionServer 2288H v5 servers with two head nodes for control. Nodes are split into two classes depending on their primary use. There are eight ingest nodes, designed for the capture and collation of Ethernet data and its writing to disk. Each ingest node has two 3-GHz Intel Xeon Gold 6136 CPUs, a 40-GbE Ethernet adapter and an 56 Gb/s FDR Infiniband adapter. The remaining 60 nodes are used for data analysis, with each having two 2.1-GHz Intel Xeon Silver 4116 CPU and two Nvidia GTX 1080 Ti GPUs. Additionally these nodes have FDR Infiniband adapters. All nodes in the cluster are connected via an Infiniband fabric with each node hosting $8 \times 8$ TB hard drives, providing a 3.2-PB distributed file-system with 40 GB/s of write performance. The APSUSE ingest nodes were used during \mtp\, observations to record beams formed by the FBFUSE instrument. At L-band, data were recorded with 4096 frequency channels over a 856-MHz band centered at 1.284 GHz with a time resolution of 76.56 $\mu$s. At UHF the data were recorded with 4096 frequency channels over a 544-MHz band centered at 816 MHz with a time resolution of 120.47 $\mu$s.

\section*{Full-time resolution data capture pipeline}

We also used the transient buffer mode of the MeerTRAP pipeline to capture complex channelized data in order to investigate the sub-pulse structure and polarization properties of the \mtp\, single pulses. For each observation, the MJD of the first detection in the real-time pipeline was used as a reference in order to predict the time of arrival of future pulses using the best known pulse period. Then, a list of predicted MJDs were sent as a trigger to the transient buffer capture code (\url{https://github.com/ewanbarr/psrdada_cpp}) to each beamformer node. On receiving the trigger, the code uses the DM of the source to extract the complex channels around the time of the pulse accounting for the dispersion delay in each frequency channel before saving the data to disk. Each beamformer node saved data for one sub-band and all antennas used in the observation. For each pulse, around 1.5 seconds of complex channel data were stored. For every pulse, the data for each sub-band were combined and the proper complex gain correction applied per antenna and per frequency channel. The complex channels from each antenna were then added coherently, in phase, to form a beam at the phase centre of the telescope. These coherently beamformed complex channels were then used to study the polarization properties. Due to storage constraints only a small fraction of the total number of pulses from \mtp\, were captured.

\section*{Astrometry}

We compare the positions of radio sources in our combined images to those of the Rapid ASKAP Continuum Survey (RACS) \citep{mcconnell20}. We achieve this by running the {\sc pybdsf} \citep{mohan07} source finder on a joint MeerKAT image (made prior to model subtraction from all of the 2 second data), and the RACS image of the corresponding region. We forego the use of the RACS catalogue and run {\sc pydsf} directly on the RACS image in order to minimize any differences that may be introduced by the use of differing source finders. Furthermore, we select only sources that are compact in both catalogues (represented by a single point or Gaussian component) in order to minimize any resolution biases. The offsets between 254 MeerKAT sources and their matched RACS counterparts are shown in Supplementary Figure \ref{fig:astrometry}. The mean offsets in RA and Dec respectively are 0.90 ($\pm$ 0.07)$''$ and $-$0.55 ($\pm$ 0.05)$''$, where the quoted uncertainties are the standard error in the mean. The standard deviations of the offsets in RA and Dec are 1.04$''$ and 0.84$''$ respectively. 

RACS' positional accuracy is compared to that of the International Celestial Reference Frame (ICRF) version 3 \citep{charlot20}. Using RACS as an intermediate step we find negligible ($-$0.052$''$) systematic offset in RA, and a $-$0.95$''$ systematic offset in Dec (approximately 1/6th the size of the restoring beam) between the MeerKAT positions and the ICRF v3.

\section*{Location in the $P-\dot{P}$ parameter space.}

Pulsars dissipate their rotational kinetic energy through electromagnetic radiation and a wind of relativistic particles leading to a gradual increase (called spin-down) of the rotation period as they age. A combination of the spin-period and spin-down or period derivative gives an estimate of a pulsar's age and magnetic field strength and places it in the  $P-\dot{P}$ parameter space which is used to compare relative populations of pulsars and to potentially trace their evolution (see Figure~1). Over the lifetime of the pulsar, it is thought to cross over a death line into a region called the pulsar graveyard \citep{handbook} where it is no longer expected to emit in the radio. The rotational parameters of \mtp\, place it in the upper right region of the $P-\dot{P}$ parameter space, below two of the three death-lines in Figure~1. These death-lines are dependent on conditions needed for pair production in the pulsar magnetosphere. The conventional emission model for pulsars assumes the existence of a vacuum gap above the polar cap of a neutron star. In order to sustain pair production, the potential difference across this gap must be sufficiently large and this is no longer possible beyond this death line. As a result, pair production and consequently, radio emission ceases. 

Equation 9 from CR93 \citep{CR93}, illustrated by the solid line in Figure~1, represents a model for very curved or twisted magnetic field lines with curvature radius of the magnetic field line, $r_{c} \sim R = 10^{6}$~cm comparable to the radius $R$ of the neutron star. It also suggests a polar cap size much smaller than that of the pure dipole field case. \mtp\ lies well beyond this death-line implying that this model is not viable.

The model for curvature radiation from the vacuum gap (represented by Equation 4 in \citep{ZHM2000} and illustrated by the dot-dashed line in Figure~1), utilizes a multipole magnetic field configuration and relativistic frame-dragging effects to achieve the necessary potential difference for continued pair production. \mtp\, is also located beyond this death-line implying that an alternative model is required to explain the radio emission.

The space-charge-limited flow (SCLF) model death line for curvature radiation (calculated using Equation 9 in \citep{ZHM2000} and illustrated by the dashed line in Figure~1) assumes a multipole magnetic field and lies just below the location of \mtp\, indicating that this model is potentially feasible. 
Additionally, it has been proposed \citep{ZTZ+17} that different equations of state of a neutron star possibly affect the death-line, thereby leading to heavier radio pulsars surviving beyond the standard death line.

For the simplest assumption of a dipolar magnetic field configuration, the minimum dipole magnetic field strength $B$ at the surface of a canonical pulsar is,

\begin{equation}
    \Bigg(\frac{B}{\rm{Gauss}}\Bigg) > 3.2 \times 10^{19}\, \Bigg( \frac{P\dot{P}}{\rm{s}}\Bigg)^{1/2},
\end{equation}

\noindent which corresponds to $1.4\times 10^{14}$~G for \mtp. This value is similar to the bulk of the magnetar population in the $P-\dot{P}$ parameter space (calculated using this same standard formula). Assuming that the source was born with an initial period much shorter than its current period, a simplified estimation  of the approximate, or characteristic age, $\tau$ can be made as,

\begin{equation}
    \tau = \frac{P}{2\dot{P}} ,
\end{equation}

\noindent which is equivalent to an age of 5.3~Myr for \mtp.
A magnetar with a spin period of $\sim10$~s and a braking index $n=3$, would evolve to the spin period of \mtp\, in $\sim5$~Myr. We note that even if $n$ were as high as 5, it still gives an age of $\sim2$~Myr.

\section*{Average Pulse Profiles and Pulse Modulation}

In Supplementary Figure \ref{fig:modulation} we show the average pulse profiles formed from the observations taken on 2021-04-02 when we observed with both the UHF and L-band separated by less than 2 hours to be able to compare the behaviour quasi-simultaneously between the two frequencies. The profiles are similar at the two frequencies with a smooth leading edge rising up to the peak of the profile and then a second component apparent on the trailing edge. The UHF profile is less smooth at the peak and has a more distinct trailing component which may be to do with the increased modulation (see below) in the UHF single pulses. Although the L-band data is from a single epoch, it is representative of the average pulse seen on other days. 

We measured the full-width at half maximum $W_{50}$ of both the pulses by fitting five and seven von Mises functions (the von Mises distribution resembles a Gaussian distribution, but is cyclic) to the L- and UHF-band profiles respectively, using the \textsc{psrsalsa} package \citep{psrsalsa}. This analytic, noise-free description of the profile allows a robust width measurement of the profile shape. The uncertainties on the measurements are estimated by bootstrapping, i.e. by repeatedly adding Gaussian noise with the standard deviation measured from the off-pulse region of the profile, over multiple iterations. The quoted uncertainty is the standard deviation of the iterations. We find that the $W_{50}$ widths of the two pulses are identical to within the errors and also to within 0.3\% of the pulse width indicating that between the two bands there is no evidence for radius-to-frequency mapping. It is interesting to note that the $W_{50}$ point lies just about the point where the trailing feature is present indicating that too is apparently not evolving dramatically with frequency. 

As can be seen in Figure~1 there are a number of different types of pulse shapes that are seen from \mtp. To try and quantify the variability we calculated the modulation index and standard deviation as a function of pulse phase from a series of pulses (see \citep{wes06} for the definition of these terms). We again used the UHF and L-band data from 2021-04-02 and the data set was the same as that used for the pulsar timing. We formed a fully-frequency averaged stack of single pulses for each of the 30 minute sessions, two each at the two frequencies, using the ephemeris given in Table~1. RFI mitigation was done before frequency averaging and was the same as described for the timing analysis above except that in some cases the pulses were corrupted too much by the RFI and these pulses were removed. We then combined the two 30-minute stacks to form a single stack for each frequency, resulting in about 45 pulses in each case. The single pulses were recorded with 65536 phase bins across a pulse period to give a time resolution of approximately 1.15 ms. The subsequent analysis was all undertaken using the \textsc{psrsalsa} package.  The off-pulse baseline was removed from each single pulse, and in some cases there was a slope in the baseline across the pulse and this was removed for doing a linear fit to the off-pulse region. The data were then gated to the section around the pulse profile as seen in Supplementary Figure \ref{fig:modulation}.

There are stark differences between the modulation properties at the two frequencies, which were observed less than 2 hours apart. We note that a similar analysis that used all the L-band pulses over the 6 epochs in 2021 gives a result very similar to the one obtained from this single day L-band data set. We only have this single epoch of UHF data, but given the stability of the L-band properties it is likely that this is representative of the UHF data and even if it is not, no similar outlier is seen in all the previous L-band observations, which still makes this difference very interesting. We note that the S/N at UHF is significantly higher than the L-band data and this is supported by the greater flux measured in the imaging data at UHF (weighted mean of 169.3~$\pm$~14 mJy beam$^{-1}$, compared to 89.3~$\pm$~2.7 mJy beam$^{-1}$ at L-band). The modulation and standard deviation at L-band are reminiscent of many other pulsars, with the former peaking at the pulse edges and the latter somewhat tracing the intensity and neither being particularly high. This correlates well with the very low pulse jitter discussed in the timing section and confirms that despite the fact that the individual pulses are radically different, they form a stable average after a surprisingly small number of pulses. In contrast the UHF data show very strong modulation with parts of the leading edge exceeding a modulation index of one and a clear dip where the transition to the trailing component happens in the average pulse profile. We also see that the trailing component is much more highly modulated than it is at L-band. The standard deviation is also very high and traces the average profile indicating that the overall pulse-to-pulse flux is changing significantly. The imaging data also shows a greater standard deviation in flux between pulses at UHF (15 mJy) than at L-band (5 mJy) confirming the phase resolved analysis. 

Observations of the 23-second pulsar \citep{TBC+18} at 350~MHz showed significant modulation in the pulse profile, that had not been seen at other frequencies, which was particularly evident in the leading component. However, these data were not contemporaneous and the profile was different than that seen at higher and lower frequencies \citep{AWB+21}.

\section*{Polarimetric Calibration}

The complex channelized data for every pulse from \mtp\, were processed using a custom-made pipeline to generate beamformed, full-Stokes dynamic spectra (see ``Voltage data capture pipeline" for more details). Then, we calibrated the polarization using the methodology given below. We assume that any leakage between the two polarization hands affects only Stokes V (defined as V=LL-RR using the PSR IEEE convention78). We also assume that the delay between the two polarization hands affects only Stokes Q and U. The calibration we apply ignores second-order effects. We perform a brute-force search for the rotation measure that maximizes the linear polarization fraction by using the \textsc{psrchive} tool \texttt{rmfit}~\citep{Hotan+2004}. With \texttt{rmfit} we select a range of rotation measures to search, in a number of trial steps. The delay between the polarization hands approximately manifests as an offset from the true rotation measure of the source, assuming the delay is frequency-independent. While this method provides a correct rotation measure, it still does not calibrate the polarization position angle (PPA) to its absolute value. The complex gain solutions that are computed by the Science Data Processor pipeline, also account for the phase offset and delays between the two dipoles of the receiver and the phase offsets and delays with respect to the sky~\citep{SJK+21}. Since the data for each single pulse were only 1.5~seconds long, we ignored the phase offsets because of the rotation of the feed with respect to the sky. In order to confirm whether we have absolute PPA calibration, we also captured complex voltage data for a very well-known, bright pulsar PSR~J0437$-$4715 whose polarization is very well measured at 1.4~GHz. We processed the data for PSR~J0437-4715 with same pipeline and corrected the full-Stokes data for rotation measure with the best known RM value taken from~\citep{OB+14}. Then we compared the resulting polarization of PSR~J0437$-$4715 with a fully calibrated profile taken from the publicly available EPN pulsar database (\url{http://www.epta.eu.org/epndb/}) \citep{dai2015}. 
% After a thorough cross-check, we realized that there was a mismatch in the sign convention we were using for our polarization. The solution was to change the sign of Stokes-V and exchange the values of Stokes-Q and Stokes-U in our data to make it fully consistent with the best known calibrated profile of PSR~J0437$-$4715. Using that as a reference, we applied this correction to all our data for \mtp. 
Extended Data Figure 6 shows examples of fully calibrated single pulses of \mtp. It is clear from the 1284~GHz pulses that the PPA follows a S-shaped sweep akin to PPAs observed for canonical radio pulsars. This is more obvious in the phased resolved PPA histogram shown in Supplementary Figure \ref{fig:pahist}.

\section*{Rotating Vector Model fits}

The Rotating Vector Model (RVM; \citep{rc69a}) provides a geometric interpretation of the variation of the PPA with rotational phase. To increase the sensitivity 11 consecutive pulses were summed together. The resulting polarimetric profile is shown in Supplementary Figure \ref{fig:rvmfit}. A rapid swing in PPA is shown near the peak of the profile, with the steepest part of the swing coinciding with the transition between the two main components of the profile. There is a distinct dip in the degree of linear polarization. Although such a dip can in principle be the result of mixture of orthogonal polarization modes, the same orthogonal mode appears to dominate in this pulse longitude region for all the analyzed pulses (see the PPA distribution in Supplementary Figure \ref{fig:pahist}).

The PPA swing as observed in individual pulses is variable, especially where the degree of linear polarization is low. Nevertheless, the steep swing is a stable feature. The PPA swing of the 11 summed pulses was fitted with the RVM following the methodology described in \citep{rwj15}. The best fit (red line in the bottom panel of Supplementary Figure \ref{fig:rvmfit}) is relatively poor (the reduced $\chi^2$ is 16), with significant deviations from what can be modelled with the RVM. The inability of the RVM to reproduce the observed PPA swing accurately means that the model parameters are poorly constrained. The magnetic inclination angle (with respect to the rotation axis) is unconstrained. What can be constrained is that the line of sight passes very close to the magnetic axis, with an impact parameter $\beta \lesssim 0.2^{\circ}$.

A small impact parameter is expected for a slowly rotating pulsar, as it will have an extremely large light cylinder, and a correspondingly small magnetic pole defined by open magnetic field lines on which particle acceleration and the production of radio emission is expected. The PPA swing therefore suggests that the observed radio emission originates close to the magnetic pole of the neutron star.

\section*{Sub-pulse width evolution}
On top of the well-defined quasi-periodic behaviour and substructure shape, multiple pulses show a complex evolution in time. An example of such a pulse can be seen in the 'quasi-periodic' panel of Figure~2. Here the pulse starts with relatively fast oscillations and the oscillations appear to slow down closer to the pulse peak. Interestingly, most quasi-periodic pulses show such `driven' oscillation suggesting some sort of a `charge and discharge' mechanism. Such behaviour may be too complex to be easily identifiable within a single ACF, which in this case is more sensitive to the shorter period.
In order to investigate the evolution of the components and their quasi-periodicity we perform a continuous wavelet transform (CWT) on pulses detected with the APSUSE instrument as those data possess a finer time resolution (76~$\mu$s). This approach allows us to analyze our signal in both the time and frequency domains. The CWT of signal $x(t)$ can be computed using the equation,

\begin{equation}
    \mathrm{CWT\left(a, b\right)} = \frac{1}{\sqrt{a}} \int_{-\inf}^{+\inf} x\left(t\right)\psi^{*}\left(\frac{t - b}{a}\right)\mathrm{dt},
\end{equation}

\noindent where $\psi(t)$ is the mother wavelet. The wavelet can then be stretched or compressed by a scale factor $a$ and shifted by the translational parameter $b$. By varying the scale factor, the wavelet transform can be sensitive to features with different sizes. By applying the shift, the transform can localize the signal details in time.
In our analysis we use two wavelets to extract different features from the pulse profiles. The first order derivative of the Gaussian wavelet, later referred to as the G1 wavelet, at a time $t$ given by
\begin{equation}
    \psi_G(t) = C_G t \exp\left({-t^2}\right) ,
\end{equation}
\noindent is used to detect sharp changes in the pulse profiles. By using a wavelet best suited to designing sharp transitions in the signal, we can obtain the start and end times of the quasi-periodic emission and derive the widths of individual components. To better understand the evolution of the quasi-period, we detect individual components and localize them in time. We can distinguish separate sub-pulses by using the Ricker wavelet of the form

\begin{equation}
    \psi_R(t) = \frac{2}{\sqrt{3} \pi ^ {\frac{1}{4}}}  \left( 1 - t^2 \right) \exp\left({\frac{-t^2}{2}}\right)
\end{equation}

Supplementary Figure \ref{fig:wavelet} shows an example wavelet transform analysis performed for one of the pulses with a strong quasi-periodic behaviour. The de-dispersed signal at a time resolution of $76~\mathrm{\mu}$s is first convolved with a window of size 64 samples. This reduces the amount of unwanted noise and makes the structures of interest more prominent, without negatively impacting the time resolution. The convolved signal is then high-pass filtered to remove the overall pulse envelope. In this case, we use an ideal high-pass filter and fully filter-out frequencies below 5~Hz. The resulting convolved and filtered signal, shown in blue in all the panels of Supplementary Figure \ref{fig:wavelet}, is passed through the two wavelet transforms described above. In the case of the G1 wavelet the trough-to-peak transition results in a positive value of the wavelet transform at the point of the transition. Conversely, the peak-to-trough transition gives a negative value of the wavelet transform. When using the Ricker wavelet, the transform has its maximum at the centre of every peak. For both transforms we then use a simple peak finding algorithm to find and extract features of interest. The bottom row of Supplementary Figure \ref{fig:wavelet} shows the resulting peak and trough width (left panel) and peak-to-peak period (right panel) measurements. For this pulse, the sub-pulse widths and separations are relatively stable and the averaged quasi-period 82.90~ms is consistent with the value of 79.7~ms obtained using the ACF method. 

We repeat this process for a sample of the detected pulses. Of those, 39 had at least 1 clearly defined sub-pulse that we were able to measure the width of. Supplementary Figure \ref{fig:subpulsedist_subpulse-QP}a shows the distribution of 145 sub-pulse widths that we have obtained with the wavelet method. The solid vertical line represents the median sub-pulse width of 49.00~ms. The dashed vertical lines represent one  median absolute deviation limits. We can see that the majority of sub-pulse widths are concentrated around the median value, with only a small number of detections beyond 100~ms. The largest width detected within our sub-pulse sample was 120.43~ms, which is close to 50\% of the FWHM at L-band reported in Table 1.

Supplementary Figure \ref{fig:subpulsedist_subpulse-QP}b shows the relationship between the quasi-period of the sub-pulse oscillations and the width of the components of the individual pulses. To reduce the complexity, only median values of component widths and their quasi-periods for every pulse are included. The solid black line represents a theoretical behaviour where the width of the sub-pulse is exactly half of the quasi-period. This describes a sub-pulse emission that spends half of the period in the `on' state, followed by half of the period in the `off' state. 
% \td{MC: Add something about bi-stable oscillator.}
For our sample of pulses, only a handful can be classified as existing in such a state, with the divergent behaviour away from this half-period line for higher quasi-periods.

\section*{Extant data}

We looked at archival synthesis images from various imaging surveys in the Southern Hemisphere. We searched for point source like emission in the closest fields in the TIFR-GMRT Sky Survey (TGSS) observed on 2016 March 15 at 150~MHz~\citep{TGSS}, the Sydney University Molonglo Sky Survey (SUMSS) observed between February 2003 and July 2007 at 843~MHz~\citep{SUMMS} and the Rapid ASKAP Continuum Survey (RACS) observed on 2019 April 30 at 887.5~MHz~\citep{RACS}. We did not find any significant emission at the best known radio position of \mtp\, in all these surveys and report a 3-sigma upper limit of 10.2~mJy, 6.6~mJy and 1.5~mJy for TGSS, SUMSS and RACS respectively. Taking the pulse-averaged flux density at 1.4~GHz as 200 $\mu$Jy and the measured spectral index of $-1.7$, i.e.\ approximately consistent with those of known rotation-powered pulsars \citep{jvk+18}, the expected flux densities would be 8.9~mJy, 0.47~mJy and 0.43~mJy for TGSS, SUMSS and RACS, respectively. Additionally, we examined the literature for pulsar time-domain surveys that had covered the \mtp\, field in the past. The field was indeed observed as part of the Parkes Multibeam Pulsar Survey (PMPS, \citep{mlc+01}) in 1999 March (pointing centre offset by about 2.2 arcmin) and there were three further PMPS pointings near it in 1999 July/August, albeit offset by about 12 - 14 arcmin. We obtained the data and searched them for single pulses and periodic emission, both blindly and by folding using the known ephemeris. We repeated the analysis without applying a high-pass filter, as was originally done in the default PMPS data processing, on the dedispersed time series. Neither the single-pulse, nor the periodicity search resulted in a detection. This is not surprising, as the data from the primary pointing are heavily affected by RFI. There was also a high-pass filter consisting of a 2-component RC time constant included in the data acquisition system which meant that longer timescale structure than 0.9 seconds would be removed from the data. However, as the pulse width is less than this, some harmonics would still get through into the spectrum.  The sensitivity penalty suffered because of this would therefore depend to a large extent on the pulse width, as well as the period. We estimate the 8-sigma flux density upper limit assuming a 1\% duty cycle to be about 0.3~mJy. This too suggests that the non-detections in archival data (both synthesis imaging and time-domain) are not unexpected.

\section*{Energy distributions}

We examine the possibility of various types of pulses corresponding to variations in the emission mechanism by computing pulse energy distributions. For this analysis we group the quasi-periodic and partially nulling pulses together, as all quasi-periodic pulses are partially nulling but not vice versa. It is not straightforward to separate these two classes without some sort of a qualitative metric which is beyond the scope of this paper. The average on-pulse energy distribution for each pulse archetype across both frequencies and all epochs is shown in Supplementary Figure \ref{fig:energydist}. We use the \textsc{psrsalsa} software suite to model pulse profile envelopes for the quasi-periodic and partially nulling pulses using a low-pass filter (see Supplementary Figure \ref{fig:profiles}) to estimate the energy a pulse would have without the dropouts in power (Panel 2 from the top in Supplementary Figure \ref{fig:energydist}). We lose $\sim40\%$ of the energy to the observed dropouts. When this loss is accounted for by the profile envelopes, the energy distribution is not very different from the distributions of the `normal' or `split-peak' pulses. 
These oscillations are likely linked to the radio emission production itself, or a periodic absorption mechanism that can suppress the radio emission. The brightest observed pulses are classified under the split-peak category and are UHF-detections which is consistent with the observed spectral index reported in Table~1. 

\section*{Optical Follow-up}
\label{sec:optical-spectrum}

High speed photometry of the 17th mag Gaia optical source near the radio position of \mtp\ was obtained on 7 and 9 January 2021, using the 1-m telescope of the South African Astronomical Observatory and the SHOC camera. On 7 January 2021, the source was observed with no filter and 5 second continuous exposures for 2.66 hr (start time BJD 245\,9222.41766). The second observation on 9 January 2021 again used no filter, and exposure times of 10 seconds continuous for 4.13 hr (start time BJD 245\,9224.37231). The second observation showed some evidence for long term variations on time scales longer than the observation length, but no photometric modulations were seen at 76 seconds.

Spectra of the same Gaia optical counterpart candidate was undertaken with SALT \citep{Buckley2006SPIE.6267E..0ZB} on 2021 February 2. Two consecutive 1200 s exposures were obtained, beginning at 01:25:53 UTC, with the Robert Stobie Spectrograph \citep{Burgh2003SPIE.4841.1463B} which used the PG900 VPH grating, covering the region 3920--6990~\AA{} at a mean resolution of 5.7~\AA{} with a 1\farcs5 slit width. The spectra were reduced using the {\sc pysalt} package \citep{Crawford2010SPIE.7737E..25C}, (\url{https://astronomers.salt.ac.za/software/pysalt-documentation/}), which corrects bias, gain, amplifier cross-talk and cosmetic defects and finally mosaics the three CCDs comprising the detector. The spectral extraction, wavelength calibration and background subtraction were all undertaken using standard {\sc iraf} routines, as was the relative flux calibration. The latter was achieved using an observation of EG21, taken on 31 January 2021. The SALT spectrum of \mtp\, was flux calibrated using an observation of the spectrophotometric flux standard, EG21, observed with the same grating setup two nights prior. Due to the inherent design of SALT, in particular the moving and variable entrance pupil, it is impossible to determine absolute fluxes of spectra. However, the relative fluxes and shape of spectra that are calibrated are reliable. In Supplementary Figure \ref{fig:spectrum} we show the spectrum of the Gaia optical counterpart candidate, which is typical of a reddened A-type star and therefore unlikely to be associated with \mtp.

\clearpage

%----------- Supplementary Figures -----------------

\begin{figure}
    \centering
    \includegraphics[width=4.0 in]{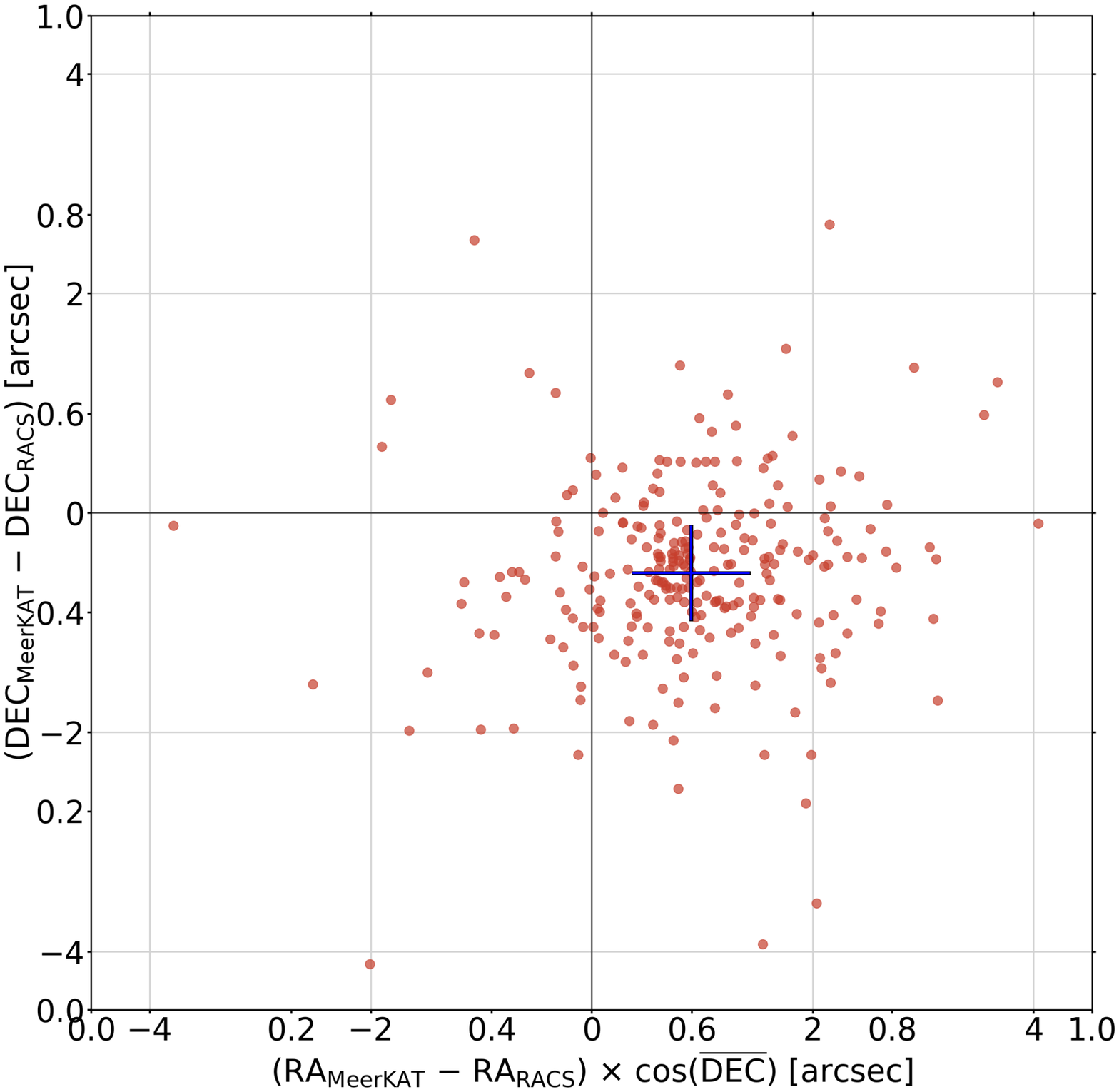}
    \caption{Offsets between the MeerKAT positions and those of the Rapid ASKAP Continuum Survey \citep{mcconnell20}. We find negligible systematic offset of $-$0.052$''$ in RA, and a $-$0.95$''$ systematic offset in Dec between the MeerKAT positions and the ICRF v3.}
    \label{fig:astrometry}
\end{figure}

\begin{figure}
    \centering
    \includegraphics[width=4.3 in]{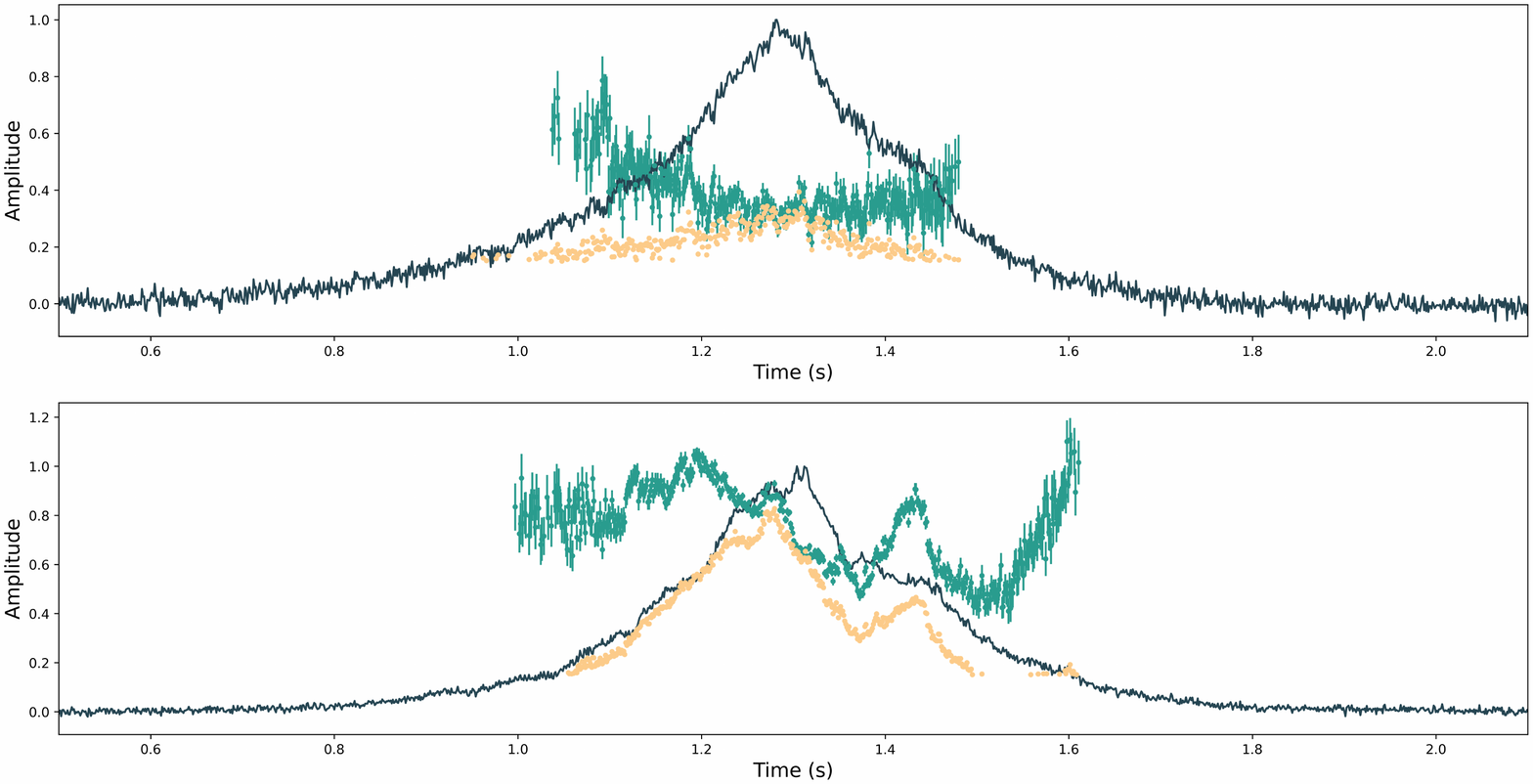}
    \caption{Pulse modulation properties of \mtp\, at UHF and L-band. A comparison between the modulation (blue), standard deviation (orange) and average pulse profiles (dark) for the L-band (top panel) and UHF (bottom panel) data taken on 2021-04-02. There are some baseline variations still in the data due to the long period and so the standard deviation is truncated at a value of 0.15 and modulation indices are only plotted when the error is less than 0.1. The profiles are scaled to have a peak amplitude of 1 and off-pulse mean of 0. }
    \label{fig:modulation}
\end{figure}

\begin{figure}
    \centering
    \includegraphics[width=4.0 in]{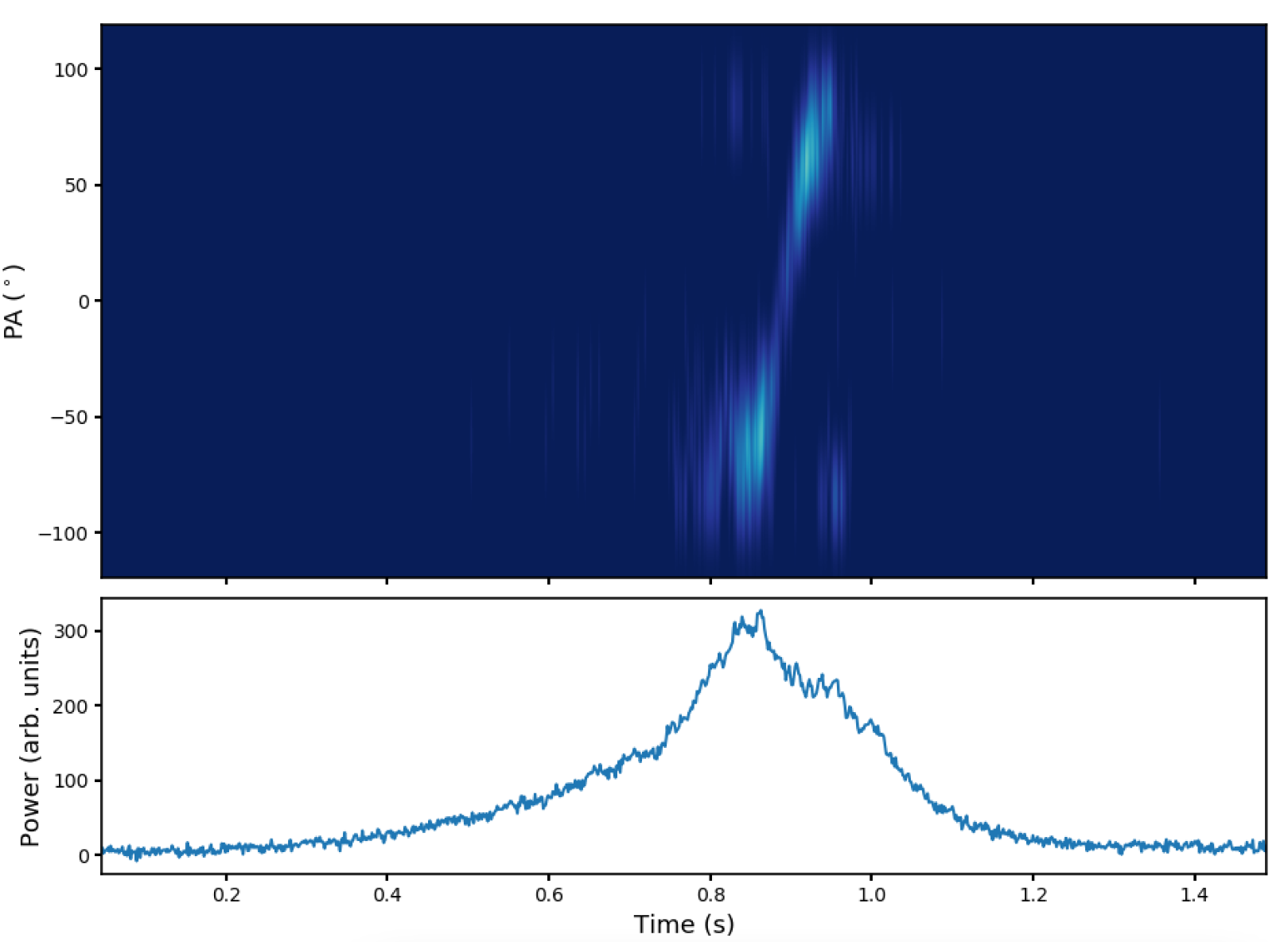}
    \caption{Phase resolved polarization position angle histogram. The top panel shows the phase resolved PPA histogram. The heat map of the PPA clearly traces out a S-shaped sweep reminiscent of the PPA swing seen in canonical radio pulsars. Bottom panel shows the average profile of the 23 pulses of \mtp\, added together.}
    \label{fig:pahist}
\end{figure}

\begin{figure}
    \centering
     \includegraphics[width=4.0 in]{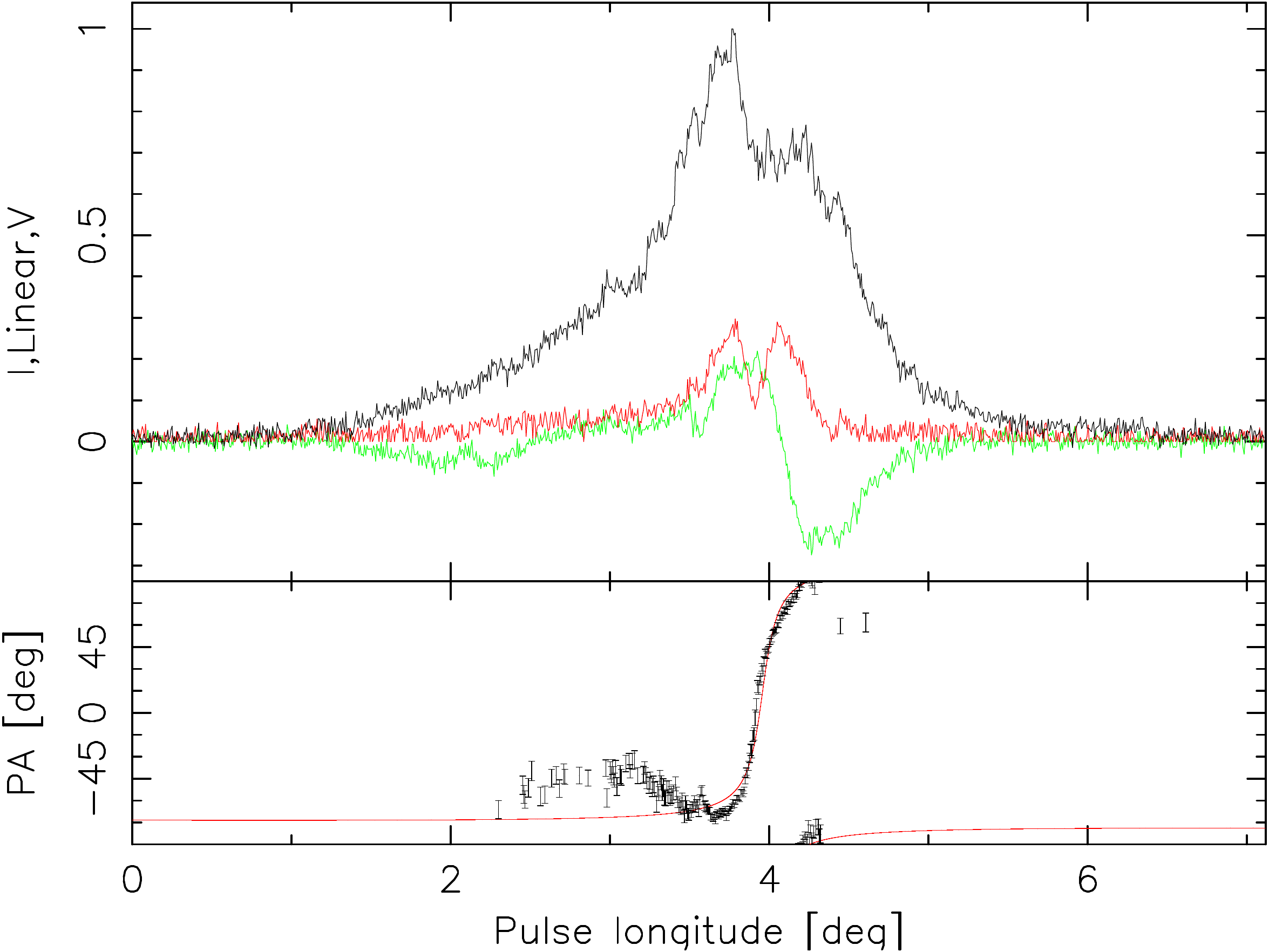}
    \caption{The polarimetric profile of \mtp\, obtained by summing 11 consecutive individual pulses recorded on the first of February 2021. Top panel: The normalized Stokes $I$ profile, the degree of linear polarization and Stokes $V$ are shown as the black, red and green curves respectively. Bottom panel: The position angle (points with error bars) as detected with $2\sigma$ confidence and the best fit RVM (red line).}
    \label{fig:rvmfit}
\end{figure}

\begin{figure}
    \centering
    \includegraphics[width=4.7 in]{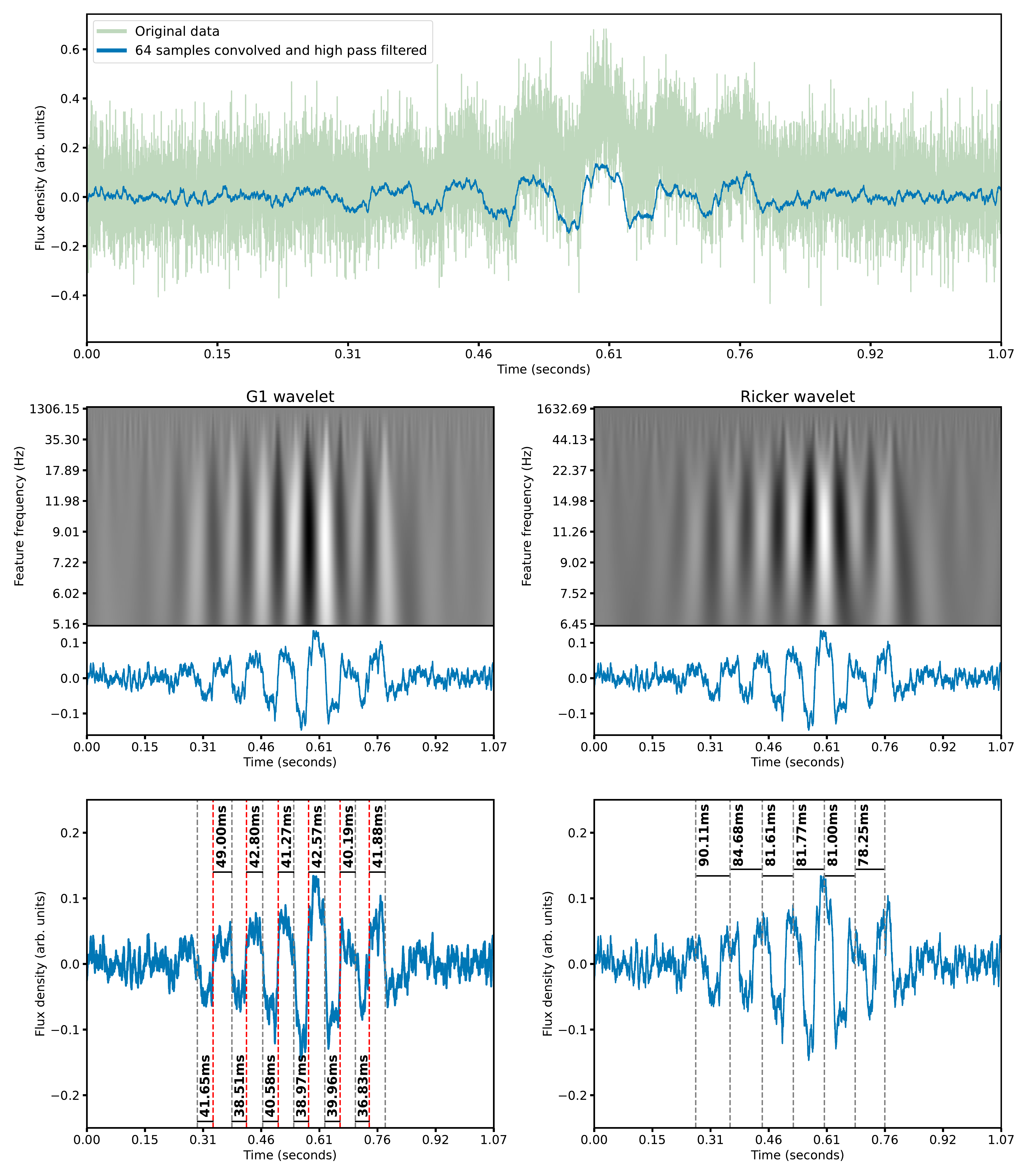}
    \caption{Example wavelet transform analysis for one of the pulses exhibiting a strong quasi-periodic behaviour. Two wavelets were used to estimate the emission properties. The first order derivative of the Gaussian wavelet was used to estimate the positions of peak-to-trough transitions, which provided an estimate of the individual sub-pulse widths and their evolution with time. The Ricker wavelet was used to determine the distance between consecutive sup-pulse peaks, providing independent and consistent estimates of quasi-periods to those obtained with the ACF method.}
    \label{fig:wavelet}
\end{figure}

\begin{figure}
    \centering
    \includegraphics[width=4.3 in]{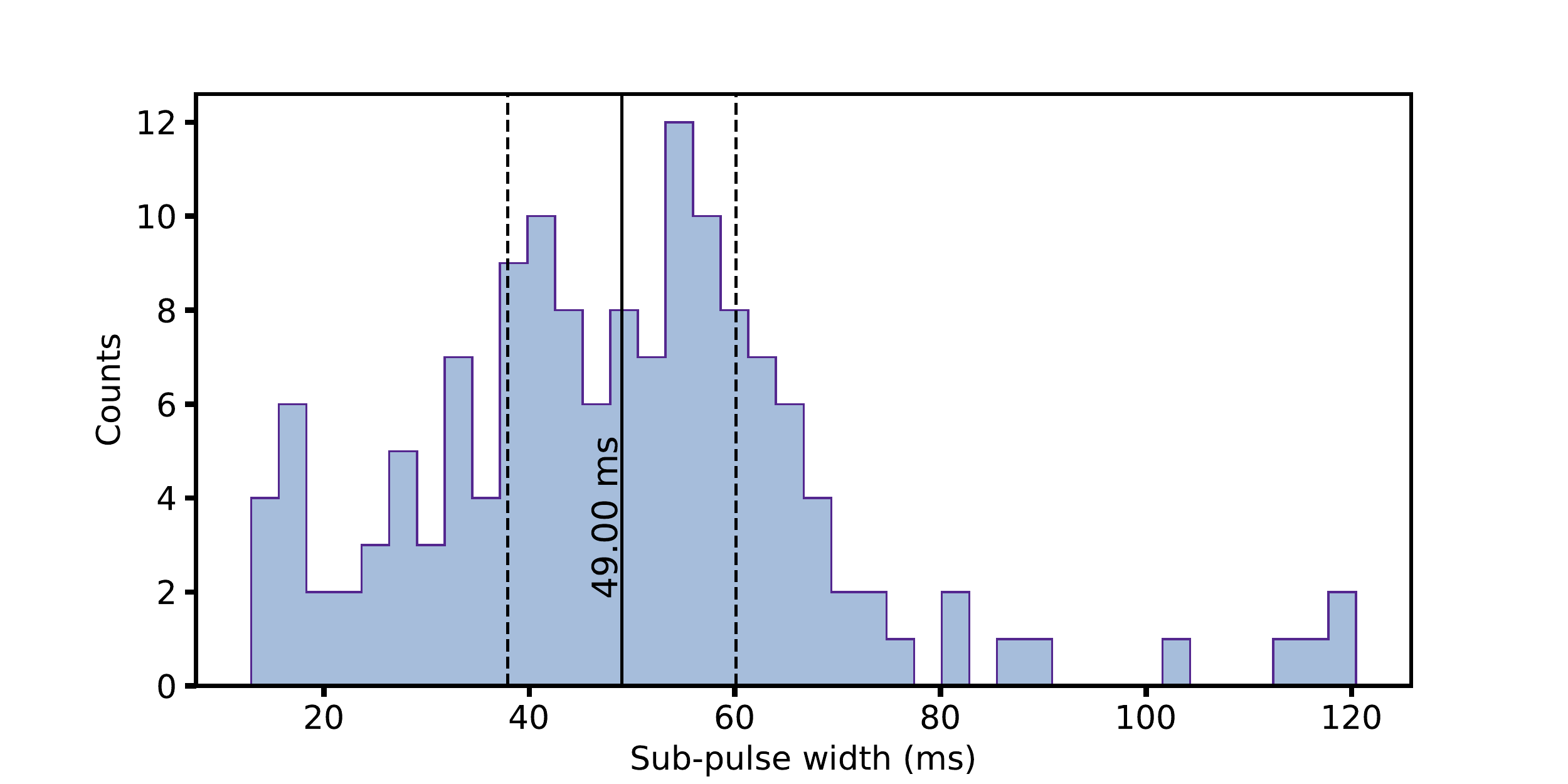}\\
    \includegraphics[width=4.3 in]{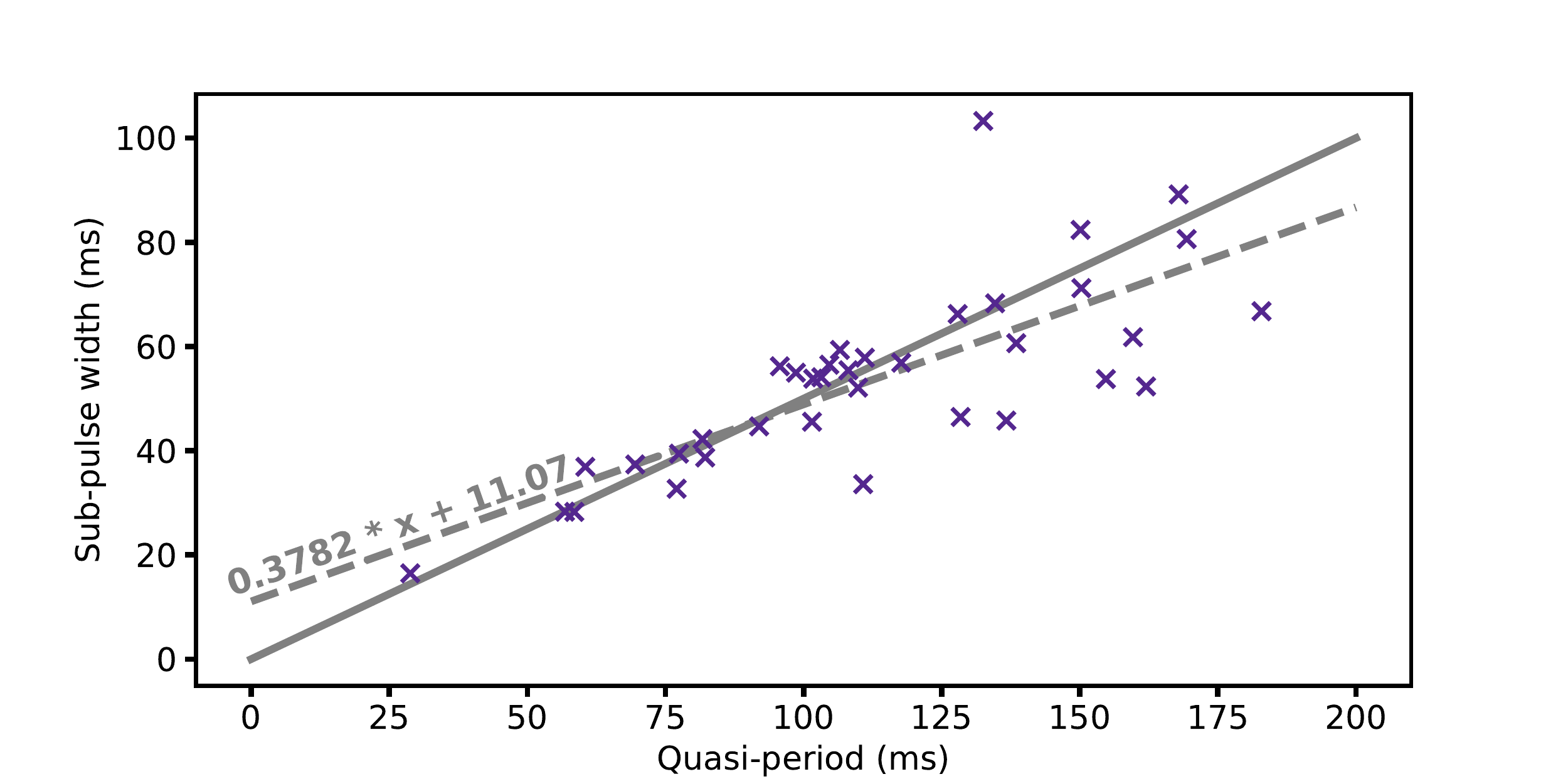}
    \caption{Sub-pulse width distribution and the quasi-period -- sub-pulse width relationship. \textit{Top panel}: Sub-pulse width distributions for widths estimated using the wavelet method. The solid vertical line represents a median sub-pulse width of 49.00~ms. \textit{Bottom panel}: Quasi-period -- sub-pulse width relationship. The solid grey line represents a behaviour where the sub-pulses widths are half of their quasi-period, which is a case in e.g. sinusoidal oscillations. The dashed line represents a least squares fit, which shows a deviation from sinusoidal oscillations.}
    \label{fig:subpulsedist_subpulse-QP}
\end{figure}

\begin{figure}
    \centering
    \includegraphics[width=5 in]{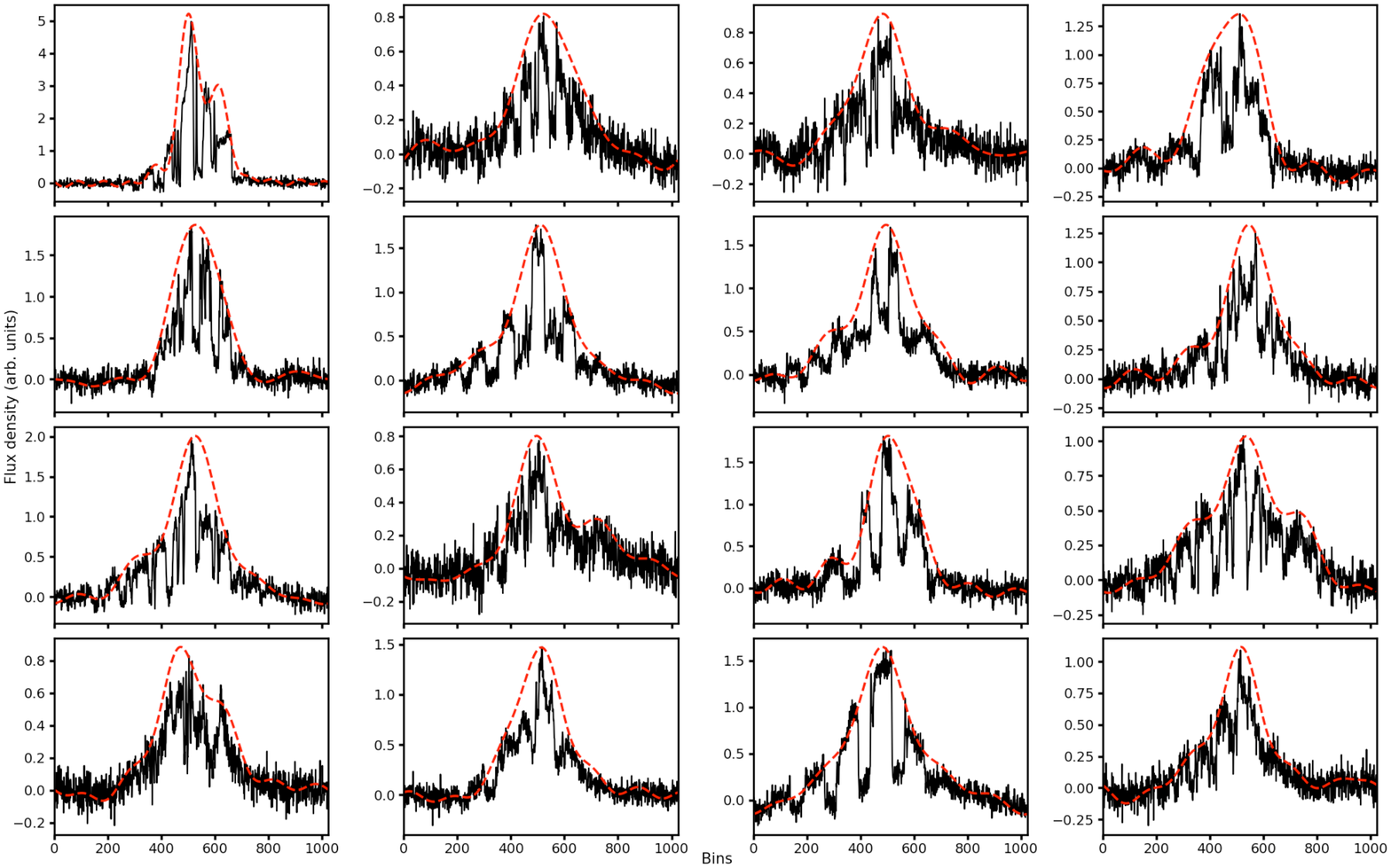}
    \caption{Sample of fitted pulse profile envelopes for quasi-periodic and partially nulling pulses. The average energies of the profiles themselves as well as modelled envelopes are represented by the solid and dashed lines respectively.}
    \label{fig:profiles}
\end{figure}

\begin{figure}
    \centering
    \includegraphics[width=4.0 in]{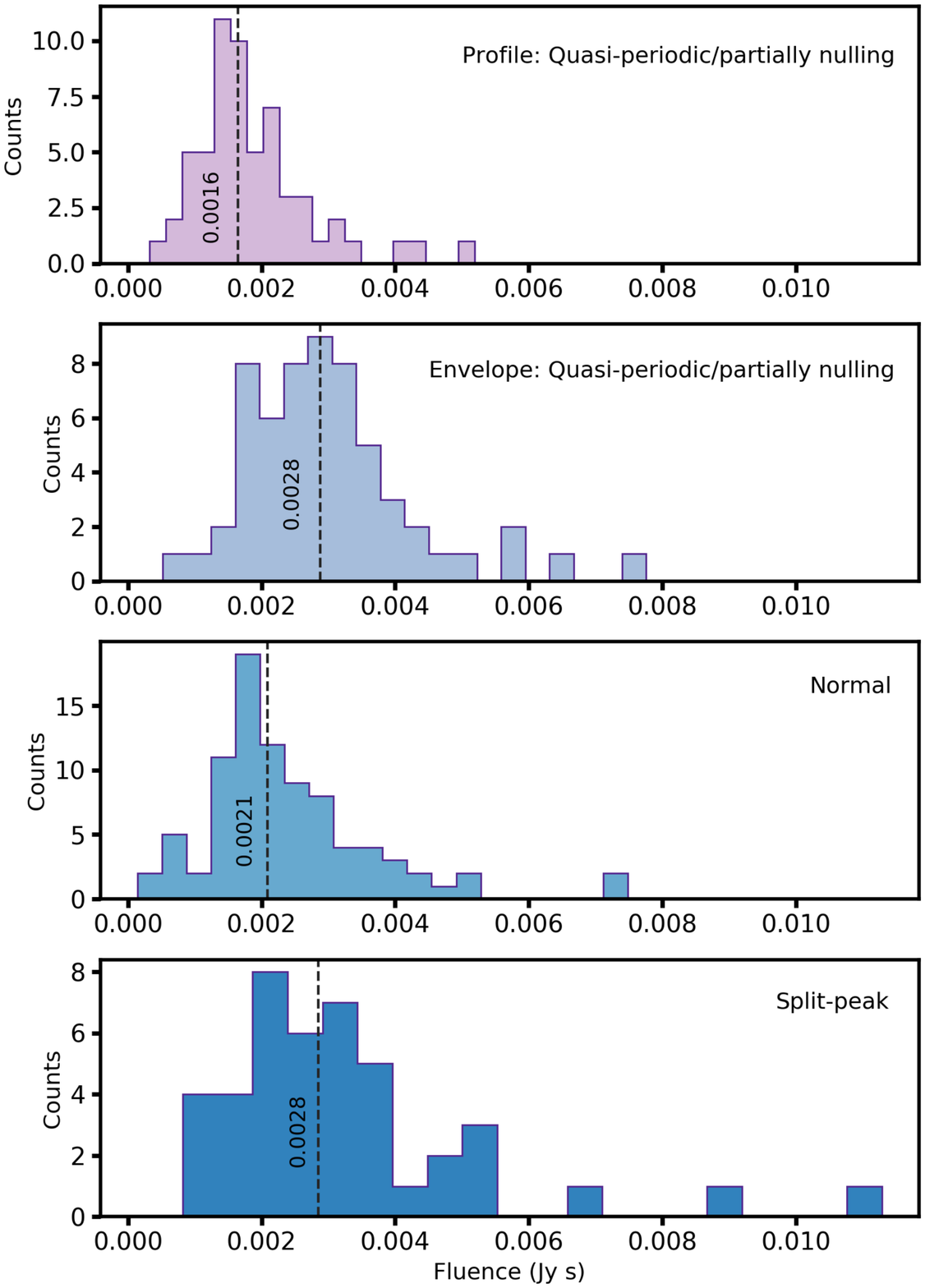}
    \caption{Pulse energy distribution of the different archetypes of pulses. $\sim40\%$ of the energy is lost to the dropouts/dips seen in the quasi-periodic and partially nulling pulses in panel 1. When this is accounted for by modelling the pulse envelope (panel 2), the resulting distribution is similar to the other distributions in panels 3 and 4. This suggests that the pulses with dropouts/dips are not drastically brighter than the other types.}
    \label{fig:energydist}
\end{figure}

\begin{figure}
    \centering
    \includegraphics[width=5 in]{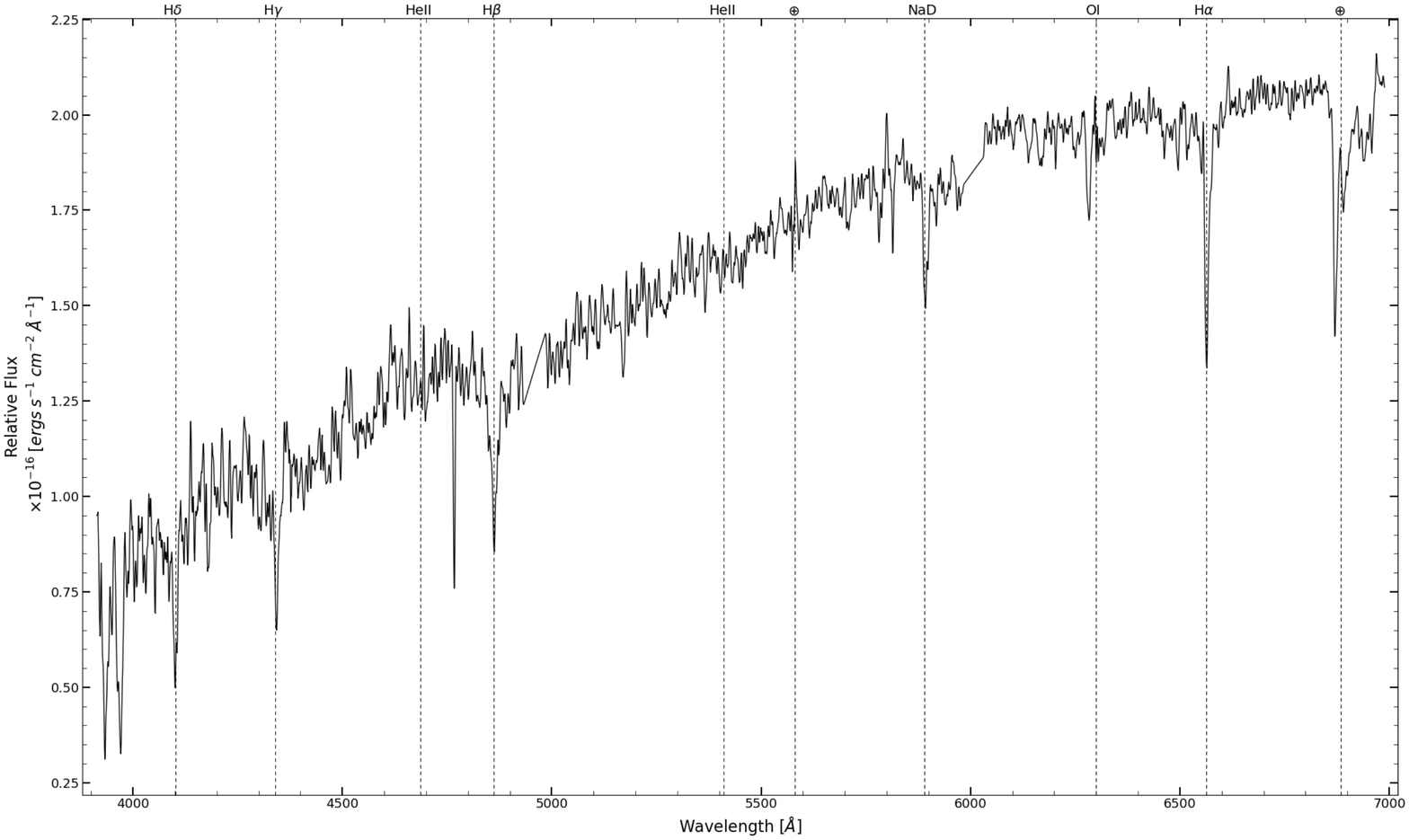}
    \caption{SALT spectrum of coincident Gaia source, ruled out as the likely optical counterpart candidate. See text for details.}
    \label{fig:spectrum}
\end{figure}

\end{document}